\documentclass[10pt]{article}
\usepackage[a4paper, margin=1in]{geometry}
\usepackage{graphicx} % Required for inserting images
\usepackage{authblk}
\usepackage[square, numbers, sort&compress]{natbib}
\RequirePackage{graphicx}
\RequirePackage{float}
\RequirePackage{subfigure}
\RequirePackage{multirow}
\RequirePackage{multicol}
\RequirePackage{makecell}
\RequirePackage{booktabs}
\RequirePackage{amsmath}
\RequirePackage[version=4]{mhchem}
\RequirePackage{caption}
\usepackage{subcaption}
\RequirePackage{xcolor,colortbl}
\usepackage{cleveref}

\title{Effects of 3D printed capsule material on activation thin foil irradiation and counting for fusion neutron yield measurements}
\author[1]{D. Lobelo\thanks{Shared first authorship: D. Lobelo and E. Panontin have contributed in equal manner to the work and writing of this manuscript and as such share first co-authorship.}}
\author[1]{E. Panontin\thanks{Corresponding author: panontin@psfc.mit.edu}}
\author[1]{X. Wang\thanks{Current affiliation: Coastal Carolina University, Conway, SC, USA.}}
\author[2]{I. Holmes}
\author[2]{P. Raj}
\author[1]{J. Rice}
\author[1]{R.A. Tinguely}

\affil[1]{Plasma Science and Fusion Center, Massachusetts Institute of Technology, Cambridge, MA 02139}
\affil[2]{Commonwealth Fusion System, Devens, MA 01434}
\date{}

\newcommand{\LaBr}{\ce{LaBr_{3}} }
\newcommand{\LaCl}{\ce{LaCl_{3}} }
\newcommand{\gammaray}{$\gamma$-ray }
\newcommand{\gammarays}{$\gamma$-rays }

\newcommand{\bla}{\cellcolor{black}}
\newcommand{\whi}{\multicolumn{1}{|l|}{}}
\newcommand{\mlr}[1]{\multicolumn{1}{l|}{#1}}
\newcommand{\mrs}[1]{\multicolumn{1}{|r}{#1}}

\begin{document}
\maketitle

\begin{abstract}
    Activation foils are used to independently measure the time integrated neutron yield and total fusion energy produced in both inertial and magnetic confinement fusion, making them crucial in the neutron diagnostic suite. The activated foils must be remotely transported from the neutron source to the detector inside of a small capsule, which can impact both the foil irradiation and the associated activation measurement. The aim of this paper is to evaluate the performance of various activation foils and to characterize the effects of different capsule materials to inform the design choices for future systems, such as the SPARC tokamak. Through a combination of FISPACT simulations and irradiation experiments with a deuterium-tritium neutron generator, we tested several different material choices for foils, capsules, and gamma-ray spectrometers. Aluminum and copper foils are found to be suitable for a multi-foil irradiation configuration. The use of 3D-printed thermoplastic capsules reduces the number of measured decay-photon counts, yet the reduction is smaller than the associated measurement uncertainty. Finally, lanthanum-based detectors are shown to be viable alternatives to the standard high-purity germanium spectrometer, although with poorer energy resolution.
\end{abstract}

%\begin{multicols}{2}
\section{Introduction}\label{sec:intro}

\begin{figure*}[t]
    \centering
    \includegraphics[width=0.5\linewidth]{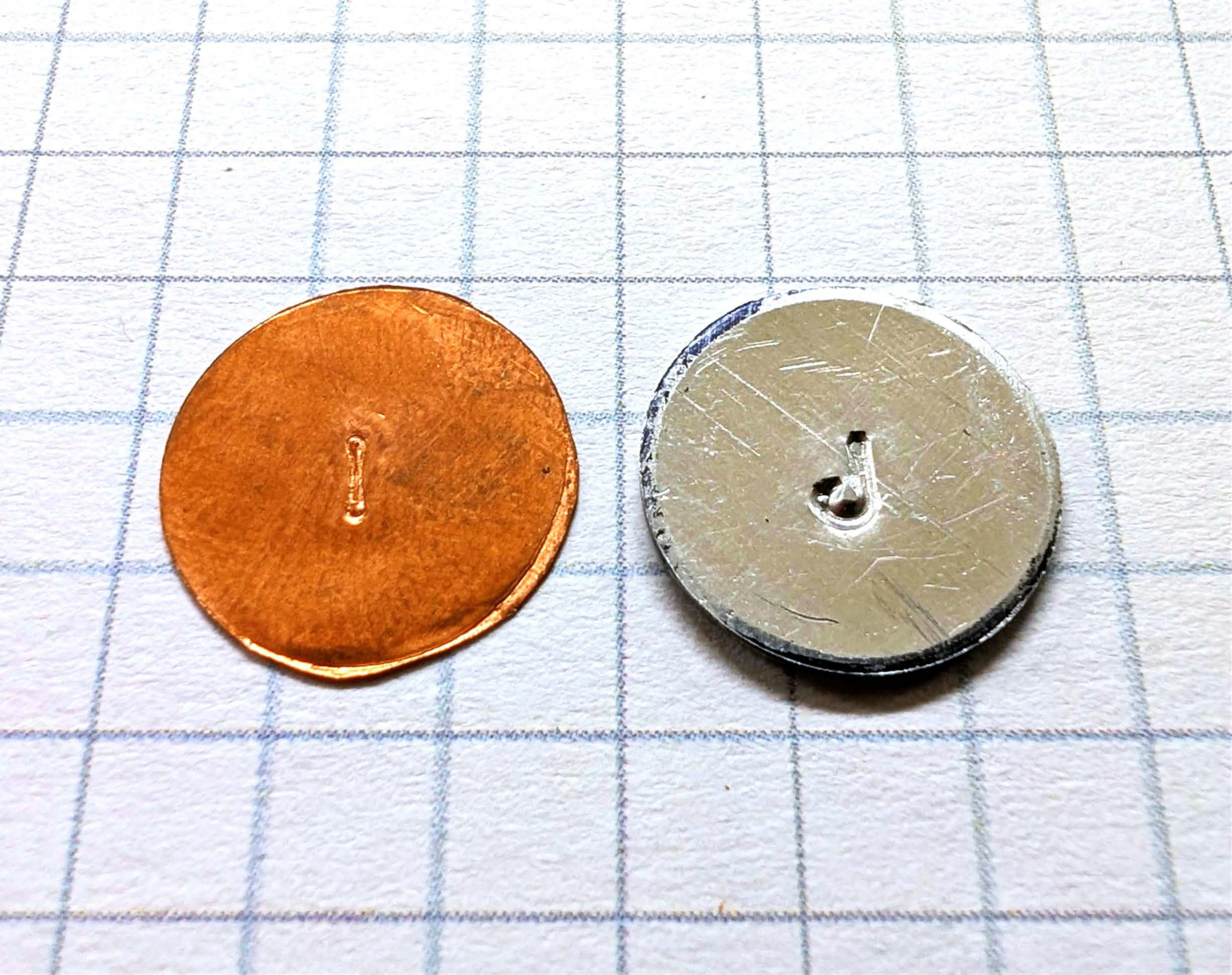}
    \caption{Foil samples, both $\sim$0.5~in ($\sim$12.7~mm) diameter, on a grid with 5~mm spacing: (left) Cu with $\sim$0.005~in ($\sim$0.13~mm) thickness, inscribed ``I'', and (right) Al with $\sim$0.03~in ($\sim$0.76~mm) thickness, inscribed ``J''.}
    \label{fig:foils}
\end{figure*}

Material activation has long been used to assess the time integrated neutron yield in past and current fusion experiments~\cite{bertalot1999, esposito1999, batistoni2018, vasilopoulou2019, barnes1990, yeamans2016, bionta2021}, and similar systems are planned for upcoming fusion devices such as ITER~\cite{krasilnikov18} and SPARC~\cite{raj2024, reinke2024}, also~\cref{app:sparc}. The working principle is that small ``foils'' with well known elemental composition, mass, and dimensions are placed close to the neutron source, in a similarly well defined location. (See~\cref{fig:foils} for sample foils used in this work).
Interactions with these neutrons transmute the nuclei in the foils into unstable isotopes, which subsequently undergo characteristic radioactive decay, emitting gamma rays with energies and half-lives specific to the foil material. %These gamma rays typically span from hundreds of keV to a few MeV.% 
Following the end of the irradiation, the activated foil is transported to a gamma ray spectrometer. 
Usually, a High Purity Germanium (HPGe) detector is used to measure the gamma spectrum of the foils, from which one can infer the neutron fluence incident on the foil and thus the total neutron yield, given a proper calibration. 
Depending on the foil materials used, foil counting can be performed using alternative detectors, such as NaI and \LaBr detectors~\cite{bionta2021, bertalot1999}.
Of particular interest for this work are lanthanum inorganic scintillators. 
They are generally cheaper than HPGe detectors and have a higher radiation hardness, but at the cost of a lower energy resolution.
Multiple foil materials, with different activation cross-sections and neutron energy thresholds, can be stacked together to infer components of the neutron energy spectrum~\cite{esposito1999}.
In turn, multi-foil experiments can provide such information as the fuel ion ratio or thermal vs nonthermal fusion fractions. 
During both irradiation and counting, the foils are housed inside a capsule for ease of transportation. Three potential capsule materials are investigated in this study: PETG (Polyethylene Terephthalate Glycol), PLA (Polylactic Acid), and PC (Polycarbonate). Multi-foil irradiations and measurements are performed in this analysis, but neutron spectrum ``unfolding'' is left for future work.

\begin{table*}[b]
\centering
\captionof{table}{\label{tab:sim:inputs} Experimental conditions considered in the FISPACT simulations performed. All simulations utilize a foil mass of 1 gram.}
\smallskip
\begin{tabular}{l r r r r r}
\toprule
 & n(DT) source rate & n(DT) flux & Irradiation time & Transfer time & Counting time\\
\midrule
SPARC & $\begin{matrix}
<5\times10^{19} \mbox{ n/s}\\
\left( P_{fus}<140 \mbox{ MW} \right)
\end{matrix}$ & $<10^{14}$ n/\ce{cm^2}/s & $<10$ s & $10$ s & $<60$ min\\
MIT & $\approx10^8$ n/s & $<10^6$ n/\ce{cm^2}/s & $>1$ hr & $90$ s & $60 - 180$ min\\
\bottomrule
\end{tabular}
\end{table*}

As will be discussed in~\cref{sec:foil:selection}, we will use copper (Cu) and aluminum (Al) foils in this work. To illustrate the activation process, we consider the case of Cu as an example. We assume a constant neutron source rate and a mono-energetic neutron flux, which is close to that emitted by a DT neutron generator. For the \(^{65}\text{Cu}(n,2n)^{64}\text{Cu}\) reaction, the amount of \ce{^{64}Cu} produced by the end of the irradiation phase can be calculated as
    \begin{equation}
        N_{\ce{^{64}Cu}}(t) = \frac{N_{\ce{^{65}Cu}}(t=0) \, \sigma \, \phi_n}{\lambda(511 \mbox{ keV})} \left( 1 - \exp\left[-\lambda(511~\mathrm{ keV}) \, t \right] \right),
        \label{eq:foil:irradiation}
    \end{equation}
where $N_{\ce{^{65}Cu}}(t=0)$ is the number of \ce{^{65}Cu} atoms present in the sample at the beginning of the irradiation. $\sigma$ is the cross-section of the \(^{65}\text{Cu}(n,2n)^{64}\text{Cu}\) reaction, which depends on the neutron energy $E_n$ and is usually measured in units of [barn]. $\phi_n$ is the flux of the neutron spectrum reaching the foil, which is usually measured in units of [1/\ce{cm^2}/s]. $\lambda(511~\mathrm{keV})$ is the rate of 511~keV $\gamma$ decays for the excited state of \ce{^{64}Cu^*}, which is related to the half-life of \ce{^{64}Cu^*} through $\lambda = \log{2} \,/\,t_\frac{1}{2}$ and $t$ is the duration of the neutron irradiation. 
If the neutron spectrum incident upon the foil is broader in energy, then we should integrate $\sigma$ and $\phi_n$ over all neutron energies; furthermore, time-evolving neutron rates must be evaluated more carefully in actual experiments.

It is worth noting that if the irradiation duration is much longer than the half-life of the unstable element, such as in the experiments discussed in~\cref{sec:neutron:irradiation}), then the foil will reach ``secular equilibrium'' for which the production rate of \ce{^{64}Cu} is balanced by its gamma decay rate, and $N_{\ce{^{64}Cu}}$ will increase only minimally for longer irradiations. Then~\cref{eq:foil:irradiation} becomes
\begin{equation}\label{eq:foil:irradiation:2}
    N_{\ce{^{64}Cu}}(t \gg t_\frac{1}{2}) \approx \frac{N_{\ce{^{65}Cu}}(t=0) \, \sigma \, \phi_n}{\lambda(511 \mathrm{ keV})},
\end{equation}
and the total number of $N_{\ce{^{64}Cu}}$ will depend only on the rate of neutrons hitting the foil.
If $t = t_1$ denotes the end of the irradiation, we can then estimate the number of gamma-rays generated between $t_2$ and $t_3$ using:
\begin{equation}\label{eq:foil:counting}
    \Delta N_\gamma (511 \mathrm{keV}) = N_{\ce{^{64}Cu}}(t_1) \, \exp{[-\lambda \, (t_2-t_1)]} \, \left(1 - \exp{[-\lambda \, (t_3-t_2)]} \right).
\end{equation}

If a foil is irradiated inside a capsule, the final number of \gammarays counted depends on any attenuation material placed around the foil. Such material could affect both the neutron flux reaching the foil and the gamma transport between the foil and the detector, reducing the final number of counts.
Such a capsule will be needed on the SPARC tokamak to transport remotely foils between the torus and the diagnostics laboratory where foil decays will be counted with dedicated \gammaray detectors.
In this paper, we characterize key components - such as foil elements, capsule material, and gamma detectors - to inform the design of activation systems, specifically that for SPARC. FISPACT simulations are used to down-select two foil elements - Al and Cu - for irradiation experiments with a DT neutron generator. From these experiments, we measure the gamma spectra of activated foils and quantify the relative attenuation introduced by various capsule materials. 
We additionally confirm the viability of a multi-foil activation setup and evaluate the performance of lanthanum bromide (\LaBr) and lanthanum chloride (\LaCl) detectors as possible alternatives to HPGe spectrometers for SPARC. 
Both \ce{^{137}Cs} and \ce{^{22}Na} gamma-ray sources are also used to supplement the foil irradiation experiments in assessing gamma-ray attenuation from the capsules. 

The rest of the paper is organized as follows:~\cref{sec:foil:selection} outlines the selection criteria and simulation results that led to the foil material choice of Al and Cu for physical testing. \Cref{sec:gamma:attenuation} provides a detailed description of the gamma-ray attenuation experiments conducted, while~\cref{sec:neutron:irradiation} focuses on the findings of foil activation experiments performed with a D-T neutron generator. A summary is provided in~\cref{sec:conclusions}. In addition,~\cref{app:sparc} describes the SPARC tokamak,~\cref{app:foil:purchase} gives specifications for the foils used in the experiments discussed in this manuscript. Finally,~\cref{app:error:normalized:counts} derives the statistical and systematic uncertainties that affect foil counting experiments.
% FISPACT simulations
\begin{figure*}[!b] 
\centering
\subfigure[]{\label{fig:fispact:decay}
\includegraphics[width=0.4\linewidth]{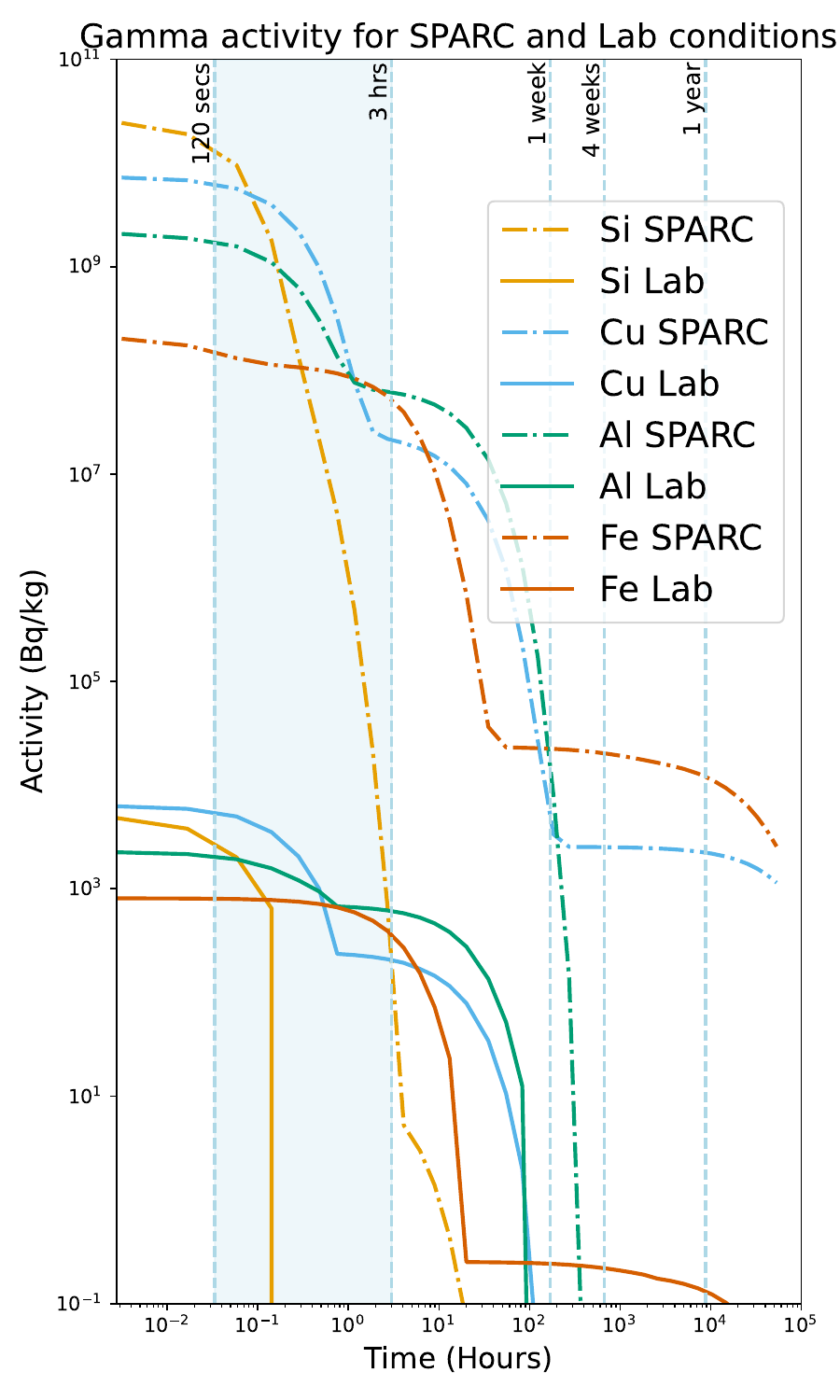}
}\subfigure[]{\label{fig:fispact:cumulative}
\includegraphics[width=0.4\linewidth]{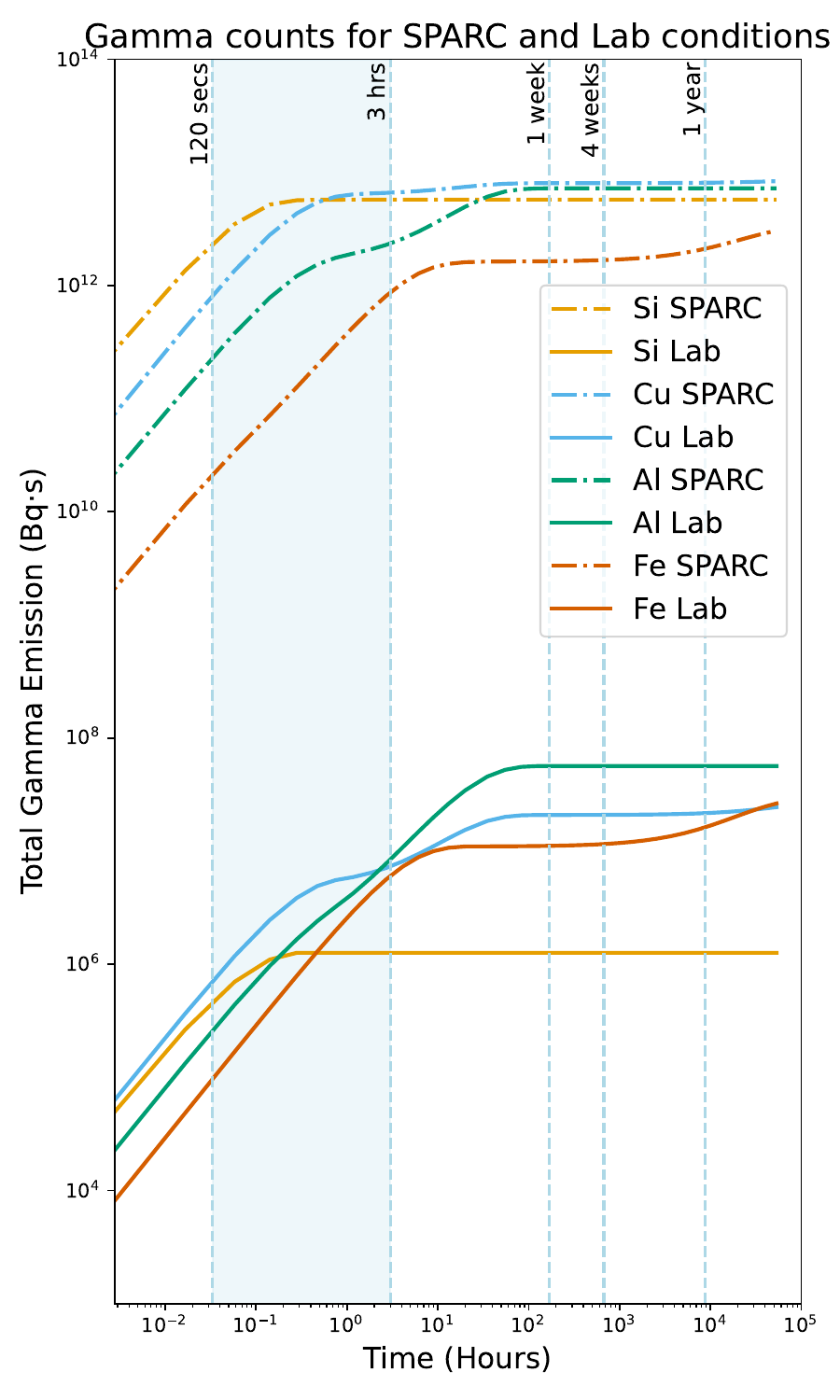}
}\\
\caption{\subref{fig:fispact:decay} Simulated gamma activity of candidate foils (Si, Cu, Al, Fe) following neutron irradiation. Two neutron irradiation cycles are presented: SPARC irradiation during a PRD and irradiation performed in the Vault Lab at MIT with a DT neutron generator. Parameters of irradiation are presented in~\cref{tab:sim:inputs}. Blue shaded background shows counting period at the MIT lab. \subref{fig:fispact:cumulative} Cumulative gamma emission from four different foil materials (Si, Cu, Al, Fe) simulated with FISPACT. Dashed lines report the gamma emission following the foil irradiation during a PRD on SPARC, while solid lines show gamma emission following irradiation in the Vault Lab at MIT.}
\label{fig:fispact}
\end{figure*}

\section{Foil Selection}\label{sec:foil:selection}
Neutronics simulations were conducted, using FISPACT-II 5.0~\cite{sublet2017} with the TENDL Nuclear Library~\cite{koning2019}, to prepare the foil irradiations discussed in~\cref{sec:neutron:irradiation}. The performance of 16 different elements were simulated using irradiation conditions typical of the DT neutron generator used in~\cref{sec:neutron:irradiation} and of SPARC. The main parameters of the irradiation scenarios considered (irradiation flux and duration, foil transfer and counting time) are listed in~\cref{tab:sim:inputs}. The neutron spectrum is assumed to be monoenergetic at 14.1MeV. The flux at the DT neutron generator has been calculated assuming a nominal production rate of $3\times10^8$n/s and foil disk placed in contact with the generator case. For SPARC operations, we considered a primary reference discharge (PRD) scenario~\cite{rodriguez2022}. During this scenario the neutron flux is expected to be $\approx10^{14}$ n/\ce{cm^2}/s, according to a high fidelity OpenMC model of the SPARC facility~\cite{wang2024, wang2025} that implements the most updated published information about the system design~\cite{raj2024, raj2026}. Each simulation consider a single irradiation cycle, then follows the time evolution of the total \gammaray activity of the foil following the end of the irradiation phase. The simulated foils are assumed to be chemically pure and have a mass of 1 gram each. This mass is close to the mass of the foils used during neutron irradiations: 1.05 g for Al and 0.56 g for Cu (see~\cref{sec:neutron:irradiation} and~\cref{tab:foil:size}). In order to plan for the experiments of~\cref{sec:neutron:irradiation}, the activity predicted by the present simulations were rescaled by the actual mass of the foils used in each run.

% Table showing isotope reactions
\begin{table*}[t]
\centering
\captionof{table}{\label{tab:foils:data:sheet} Summarizing the composition and activation characteristics of the natural Cu and Al foils tested. Reactions listed generate unstable nuclei which decay and emit gamma-rays of energy $E_\gamma$. Data from~\cite{lee2019}.}
\smallskip
\begin{tabular}{l l r r r}
\toprule
Isotope (abundance) & Reaction & $E_\gamma$ [keV] & Half-life & $\sigma$ ($E_n = 14$ MeV)\\
\midrule
\ce{^{27}Al} (100\%)      & \ce{^{27}Al}(n,p)\ce{^{27}Mg}         & 844 \& 1015    & 9.5 min & $7.31\times10^{-2}$ barn \\
      & \ce{^{27}Al}(n,$\alpha$)\ce{^{24}Na}  & 1369 \& 2754    & 15 hr & $1.23\times10^{-1}$ barn \\
\ce{^{63}Cu} (69.15\%)    & \ce{^{63}Cu}(n,2n)\ce{^{62}Cu}        & 511           & 9.7 min & $4.2\times10^{-1}$ barn \\
\ce{^{65}Cu} (30.85\%)    & \ce{^{65}Cu}(n,2n)\ce{^{64}Cu}        & 511           & 12.7 hr & $8.37\times10^{-1}$ barn \\
\bottomrule
\end{tabular}
\end{table*}

The results of the simulations of 4 elements are reported in~\cref{fig:fispact} as an example. We used them to select suitable materials to test with the DT neutron generator. The shortest half-life of a chosen foil must be sufficiently long to allow for transportation from the irradiation site to the detector. This takes 120 seconds for laboratory experiments, and it is assumed to be around 10 seconds for SPARC. It can be found that an isotope with a half-life of 8 minutes retains 84\% of its initial activity at 120 seconds, which is considered satisfactory. On the other hand, the longest half-life of the chosen foil must be short enough to allow for weekly reuse of the irradiated foil. If we assume negligible activity after 8 half-lives (0.39\% of initial activity), that means an upper bound for the half-life of $21$ hr.

\begin{table}[!b]
    \centering
    \begin{tabular}{l r r r r r r r r r}
    \multicolumn{10}{l}{\it HPGe} \\
    \midrule
              & Al   &      &      &      &      &      &      &      &      \\ \cline{2-2}
    \mlr{Cu}  & \whi & \mrs{Cu} &      &      &      &      &      &      &      \\ \cline{2-3}
    \mlr{Fe}  & \bla & \whi & \mrs{Fe}   &      &      &      &      &      &      \\ \cline{2-4}
    \mlr{In}  & \whi & \whi & \whi & \mrs{In}   &      &      &      &      &      \\ \cline{2-5}
    \mlr{Ni}  & \whi & \whi & \whi & \whi & \mrs{Ni}   &      &      &      &      \\ \cline{2-6}
    \mlr{Zn}  & \whi & \whi & \whi & \whi & \whi & \mrs{Zn}   &      &      &      \\ \cline{2-7}
    \mlr{Mg}  & \bla & \whi & \whi & \whi & \whi & \whi & \mrs{Mg}   &      &      \\ \cline{2-8}
    \mlr{Zr}  & \whi & \whi & \whi & \whi & \whi & \whi & \whi & \mrs{Zr}   &      \\ \cline{2-9}
    \mlr{Nb}  & \whi & \whi & \whi & \whi & \whi & \whi & \whi & \bla & \mrs{Nb}   \\ \cline{2-9}
              & Al   & Cu   & Fe   & In   & Ni   & Zn   & Mg   & Zr   &   \\
     & & & & & & & & & \\
    \multicolumn{10}{l}{\it\LaBr} \\
    \midrule
              & Al   &      &      &      &      &      &      &      &      \\ \cline{2-2}
    \mlr{Cu}  & \whi & \mrs{Cu} &      &      &      &      &      &      &      \\ \cline{2-3}
    \mlr{Fe}  & \bla & \whi & \mrs{Fe} &      &      &      &      &      &      \\ \cline{2-4}
    \mlr{In}  & \whi & \whi & \whi & \mrs{In} &      &      &      &      &      \\ \cline{2-5}
    \mlr{Ni}  & \bla & \whi & \whi & \whi & \mrs{Ni} &      &      &      &      \\ \cline{2-6}
    \mlr{Zn}  & \whi & \whi & \whi & \whi & \whi & \mrs{Zn} &      &      &      \\ \cline{2-7}
    \mlr{Mg}  & \bla & \whi & \whi & \whi & \bla & \whi & \mrs{Mg} &      &      \\ \cline{2-8}
    \mlr{Zr}  & \whi & \whi & \whi & \whi & \whi & \whi & \whi & \mrs{Zr} &      \\ \cline{2-9}
    \mlr{Nb}  & \whi & \whi & \whi & \whi & \whi & \whi & \whi & \bla & \mrs{Nb} \\ \cline{2-9}
              & Al   & Cu   & Fe   & In   & Ni   & Zn   & Mg   & Zr   &          \\
     & & & & & & & & & \\
    \multicolumn{10}{l}{\it\LaCl}\\
    \midrule
               & Al   &      &      &      &      &      &      &      &      \\ \cline{2-2}
    \mlr{Cu}  & \whi & \mrs{Cu} &      &      &      &      &      &      &      \\ \cline{2-3}
    \mlr{Fe}  & \bla & \whi & \mrs{Fe} &      &      &      &      &      &      \\ \cline{2-4}
    \mlr{In}  & \whi & \whi & \whi & \mrs{In} &      &      &      &      &      \\ \cline{2-5}
    \mlr{Ni}  & \bla & \whi & \whi & \whi & \mrs{Ni} &      &      &      &      \\ \cline{2-6}
    \mlr{Zn}  & \whi & \whi & \whi & \whi & \whi & \mrs{Zn} &      &      &      \\ \cline{2-7}
    \mlr{Mg}  & \bla & \whi & \whi & \whi & \bla & \whi & \mrs{Mg} &      &      \\ \cline{2-8}
    \mlr{Zr}  & \whi & \whi & \whi & \bla & \whi & \whi & \whi & \mrs{Zr} &      \\ \cline{2-9}
    \mlr{Nb}  & \whi & \whi & \whi & \bla & \whi & \whi & \whi & \bla & \mrs{Nb} \\ \cline{2-9}
              & Al   & Cu   & Fe   & In   & Ni   & Zn   & Mg   & Zr   &          \\
    \end{tabular}
    \caption{Prospective pairs for multi-foil measurements using HPGe, \LaBr,  or \LaCl  detectors. A pair of foil is white if the detector can resolve all their activation-induced peaks, black if at least one pair of peaks overlaps.}
    \label{tab:multifoil}
\end{table}

We are also interested in a large cross-section for interactions with fast-neutrons and a well-understood gamma emission spectra. The limiting factor, here, is to get enough counts during irradiations on the DT generator. In order to perform two experiments per day, foils can at most be counted for three hours. Therefore, we prefer elements that maximize the number of decay events within the three hours following the end of irradiation. A graphical comparison of the total counts expected within the three hour mark is shown in~\cref{fig:fispact:cumulative}, where the cumulative activity plot estimated by FISPACT is reported.

% Comparison of rabbit materials
\begin{table*}[t]
\centering
\captionof{table}{\label{tab:rabbit:material} Comparison of potential 3D-printed thermoplastic capsule materials for SPARC.}
\smallskip
\begin{tabular}{l r r r}
\toprule
Material & PLA & PETG & PC\\
\midrule
Benefits & High purity & Good impact resistance & Highest melting point\\
Melting point [C] & 160 & 225 & 228\\
Density [g/\ce{cm^3}] & 1.24 & 1.25 & 1.20\\
\bottomrule
\end{tabular}
\end{table*}

From~\cref{fig:fispact:decay}, silicon is one example of an element that could perform well on SPARC, but it decays too quickly for lab testing. Over the $120$ s needed to transfer the foil from the generator to the counting station, we would miss most of its \gammaray emission. On the other hand, Cu and Al exhibit a good number of total decay events, which were deemed possible to measure in the MIT laboratories. Furthermore their cross-sections favor $14.1$ MeV neutrons, so they are more sensitive to non-scattered DT neutrons. Using the results in~\cref{fig:fispact} and isotope cross-sections~\cite{lee2019}, natural Cu and Al foils were chosen for DT neutron generator experiments. \Cref{tab:foils:data:sheet} summarizes the relevant characteristics. In our experiments, we prioritized the reactions with shorter half-lives, in particular \ce{^{27}Al}(n,p)\ce{^{27}Mg}. We adopted a 1-hour duration for both irradiation and counting, which corresponds to roughly $6$ times the half-life of that reaction. This ensures that the concentration of \ce{^{27}Mg} approaches secular equilibrium at the end of the irradiation phase and we measure most of its $\gamma$ decays within the counting phase. It is worth noting that Fe could also be a good candidate for multi-foil experiments with Al or Cu. Fe is expected to yield a significant \gammaray signal that could be measured at the MIT lab and has a cross-section that favors neutrons with energies below $12$ MeV, thus complementing Al and Cu. Its only drawback is the production of a long-lived activation product: \ce{^{54}Fe}(n,p)\ce{^{54}Mn} ($t_\frac{1}{2} = 312.2$ days)~\cite{zolotarev2013, lee2019}. Even if \ce{^{54}Mn} is expected to have a low activity, it emits $835$ keV \gammaray which overlap with the Al $\gamma$ spectrum. In multi-foil experiments, we would need to conduct a detailed study of the background of the Al \gammaray spectrum to assess the systematic error introduced by \ce{^{54}Fe}(n,p)\ce{^{54}Mn}.

All of the foils considered in our simulations are expected to generate significant statistics of \gammaray decays after a single SPARC discharge. From~\cref{fig:fispact:cumulative}, even the foil with lowest activation would yield about $4$ orders of magnitude more \gammarays than Al during the tests with the DT generator. As will be discussed in~\cref{sec:neutron:irradiation}, Al peaks counted by a $1\times1$ \ce{inch^2} \LaBr cylindrical detector had a statistical uncertainty below $5$ \%. Thus the expected statistics on SPARC would be enough to provide a precise measurement of the total fusion energy produced \ce{E_{fus}}. However, there is interest in preparing multifoil measurements for SPARC, to maximize the accuracy of and the information contained in the foil measurement. For example, measuring the $844$ keV emission from Al with the $336$ keV emission from In, one can measure both DT and DD neutrons and infer the D-to-T fuel ratio. Furthermore, the (n,p) reaction of Ni and Fe has a high cross-section between $5$ and $12$ MeV and could be used to study the downscatter from the DT peak at the foils position. Finally, deploying two foils with high cross sections around $14.1$ MeV, such as Al and Zn, would increase the statistical accuracy with which \ce{E_{fus}} can be inferred with foil spectroscopy.

Ideally, we would like to select a multi-foil mixture for which the detector used for foil counting can resolve each \gammaray peak individually. 
At first order, a detector can resolve to peaks if the difference in energy is greater than the sum of the FWHM for the two peaks, i.e. if:
\begin{eqnarray}\label{eq:resolve:gamma}
    \left|E_1-E_2\right| &>& \frac{1}{2}\left(\mbox{FWHM}_1 + \mbox{FWHM}_2\right) \\
                        &>& \frac{1}{2}\left(R_1 E_1 + R_2 E_2 \nonumber\right)
\end{eqnarray}
where $R_i$ is the detector resolution for the $i$-th peak. We have experimentally measured the resolution of \LaBr and \LaCl for the gamma-ray peaks emitted by Na and Cs sources in~\cref{sec:neutron:irradiation}, and we can use the analytical fitting reported in~\cref{fig:calibration:resolution} to extrapolate this measurement between $[0.1, 3]$ MeV. 
The formula used to describe the resolution as a function of energy is $R(E) = \frac{k}{\sqrt{E}}$, where $k$ is a numerical constant. 
Even though HPGe detectors were not employed in this work, their resolution at $662$ keV is typically $0.3$ \%~\cite{szymanska2008}.

In~\cref{tab:multifoil} we evaluate which pairs of foils can be measured with HPGe, \LaBr and \LaCl detectors. 
None of the peaks emitted by the foils taken into consideration would overlap significantly with the \LaBr and \LaCl intrinsic background, which shows a peak at around $1.465$ MeV emitted by \ce{^{168}La} and \ce{^{40}K}. 
The foil pairs marked in black are those for which at least two gamma-ray peaks would overlap.
Thanks to its high energy resolution, HPGe is capable of measuring most foil combinations. 
The pairs marked in black usually share the same gamma decay, as for \ce{^{27}Al}(n,$\alpha$)\ce{^{24}Na} and \ce{^{24}Mg}(n,p)\ce{^{24}Na}, both of which emit two \gammaray at $1369$ and $2754$ keV. 
It is worth noting that the two decay channel of Cu also overlap, since they both emit $511$ keV \gammarays (see~\cref{tab:foils:data:sheet}). 
This can complicate the analysis of experimental spectra on SPARC and should be taken into account when selecting the foils to deploy during DT operations.
\LaBr and \LaCl have fewer combinations for which no peaks overlap, with \LaBr slightly outperforming \LaCl. 
For both of them, Al overlaps with Fe or Ni foils, which could potentially complicate the use of these detectors to study the energy distribution of neutrons with Al-Fe-Ni foil mixtures.
At the same time, both detectors can resolve all Al and In peaks and thus can be used to study the D-T fuel ratio.

% Rabbits
\begin{figure*}[!b] 
\centering
\subfigure[]{\label{fig:rabbit:pla}
\includegraphics[width=0.3\linewidth]{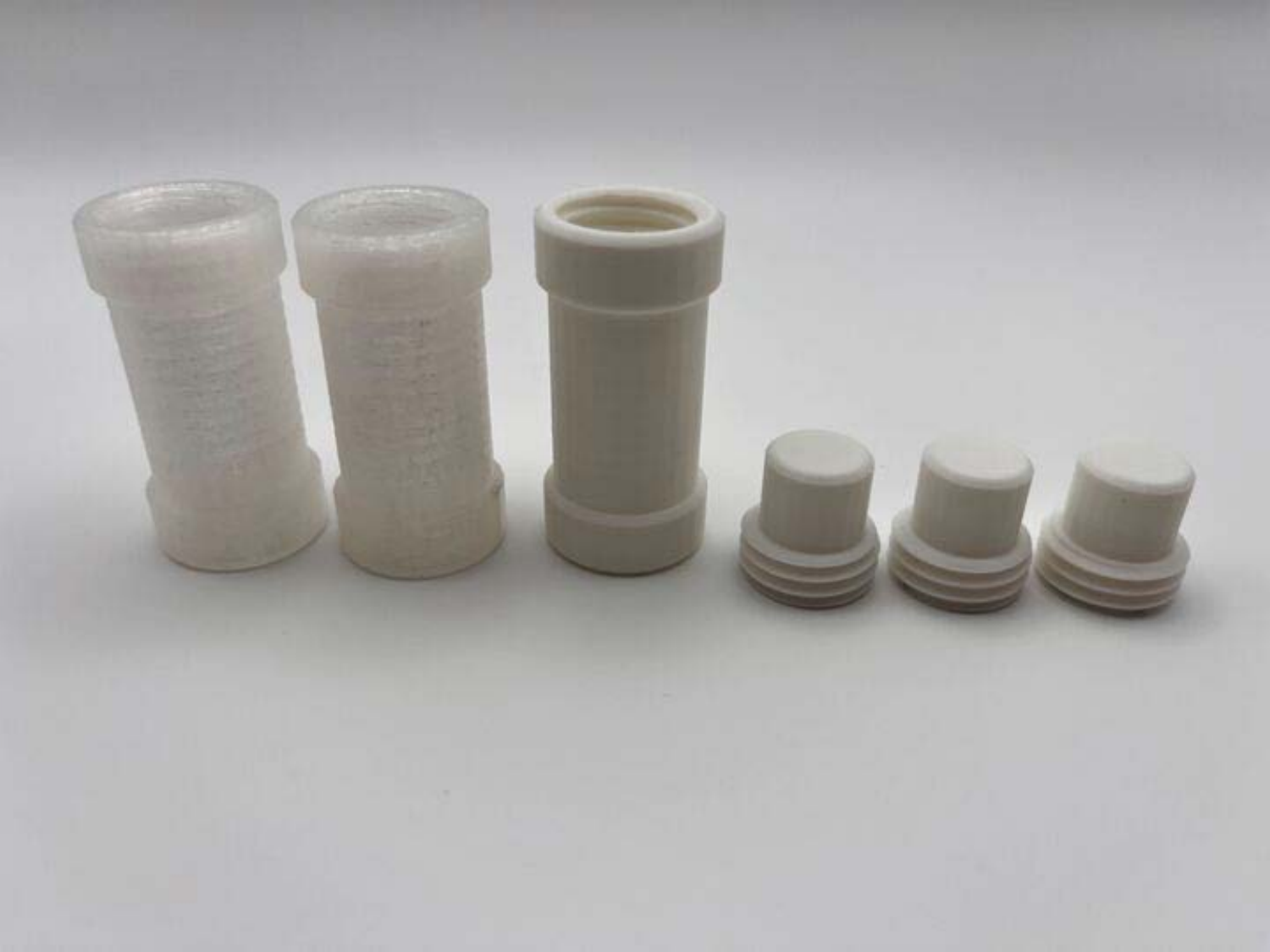}
}\subfigure[]{\label{fig:rabbit:petg}
\includegraphics[width=0.3\linewidth]{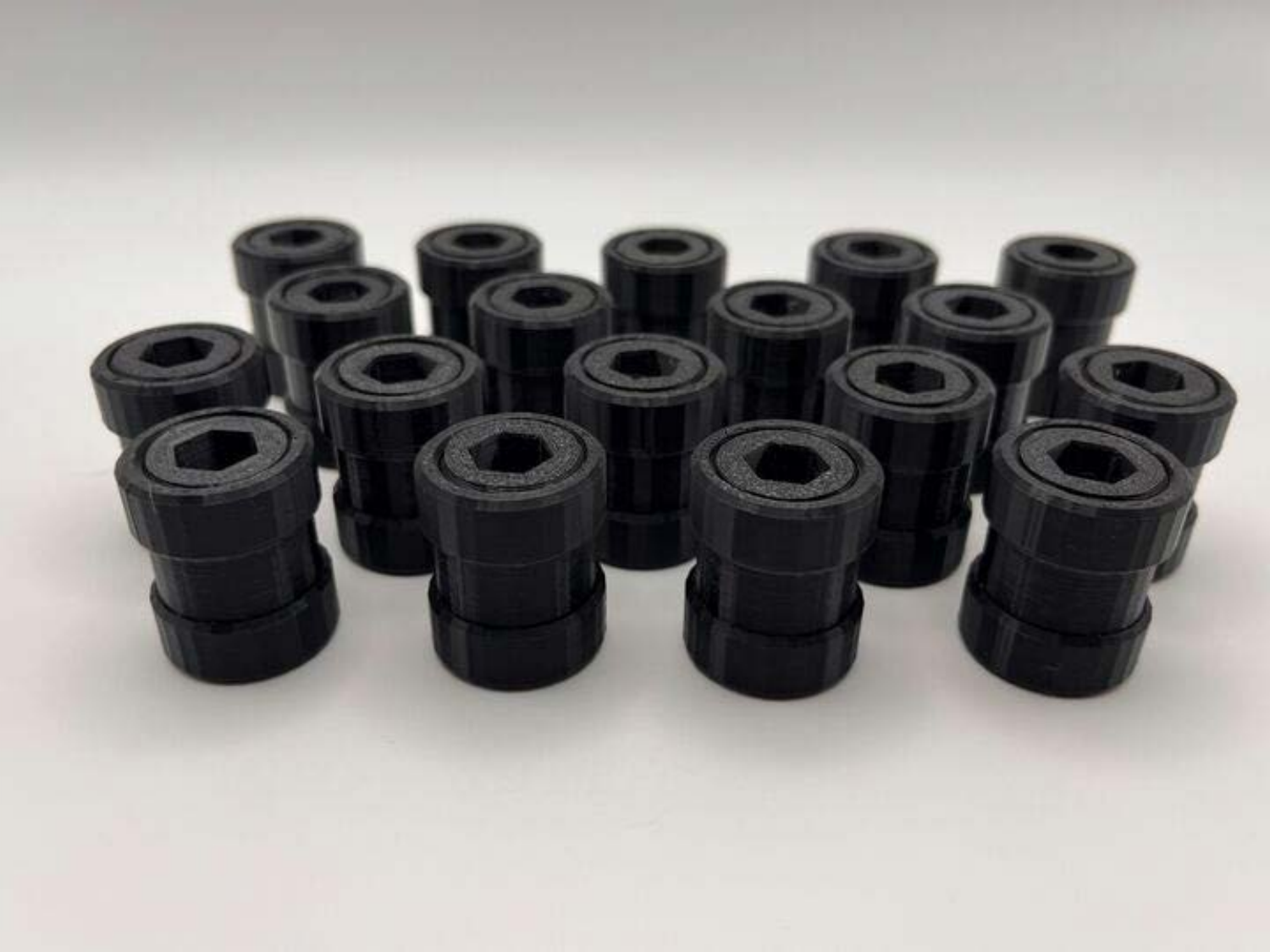}
}\subfigure[]{\label{fig:rabbit:pc}
\includegraphics[width=0.3\linewidth]{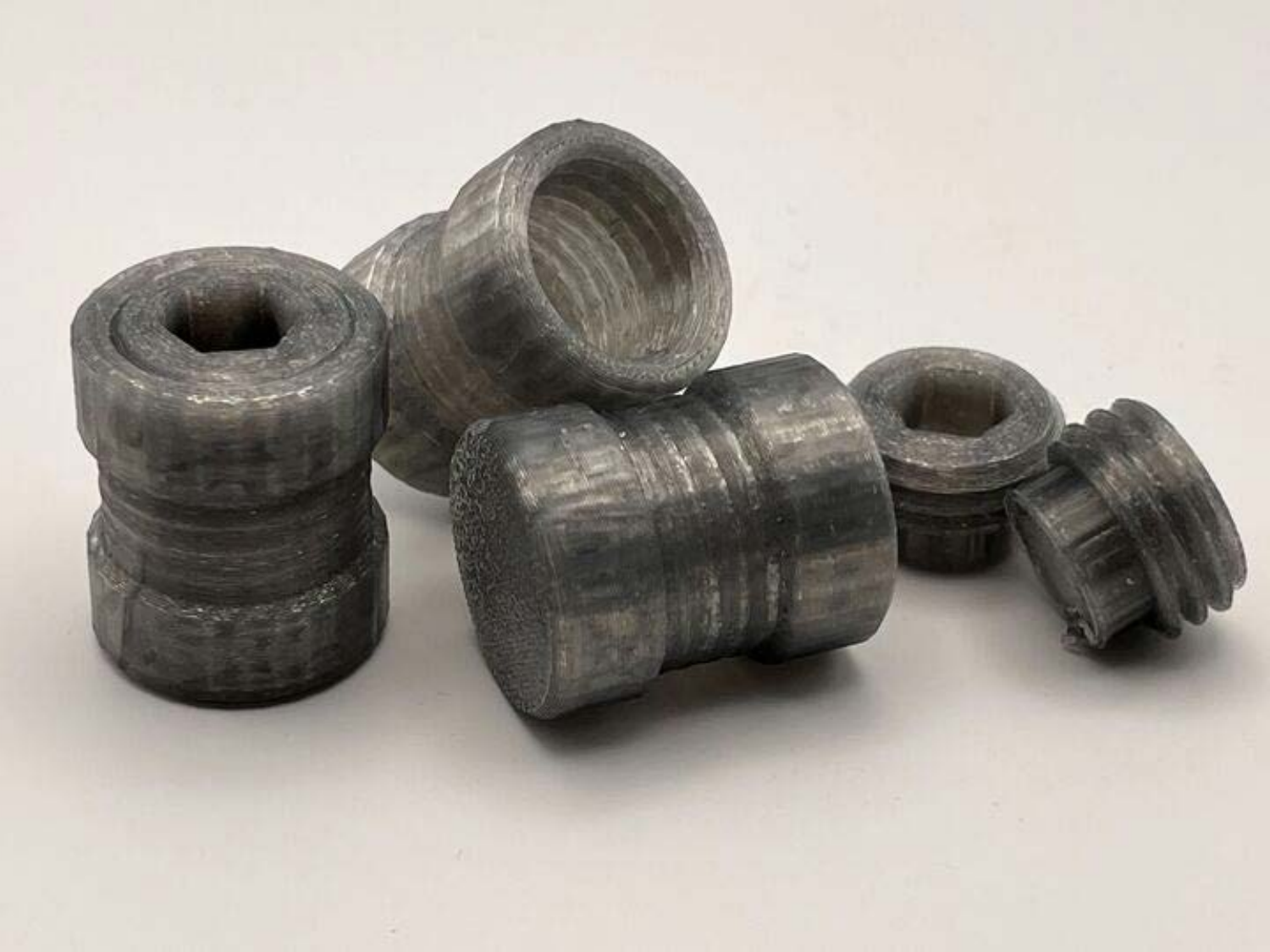}
}\\
\caption{Sample capsules made out of three potential thermoplastic capsule  materials:~\subref{fig:rabbit:pla} Polylactic Acid (PLA),~\subref{fig:rabbit:petg} Polyethylene Terephthalate Glycol (PETG),~\subref{fig:rabbit:pc} Polycarbonate (PC).}
\label{fig:rabbit}
\end{figure*}

\section{Gamma attenuation}\label{sec:gamma:attenuation}
In this section, we present the attenuation coefficient of $662$ keV gammas for three thermoplastic capsule materials: PETG (Polyethylene Terephthalate Glycol), PLA (Polylactic Acid), PC (Polycarbonate). The use of gamma sources enables quick and repeatable measurements, without the need of a long neutron irradiation and allowing for improved statistics. Moreover, this approach allows us to isolate the gamma attenuation introduced by the capsule, as the capsule could attenuate gammas \textit{and} neutrons during irradiation experiments. The combined attenuation is discussed in~\cref{sec:neutron:irradiation}. See~\cref{fig:rabbit} and~\cref{tab:rabbit:material} for photographs and details of the capsules tested.

\begin{figure*}[t] 
\centering
\includegraphics[width=0.35\linewidth]{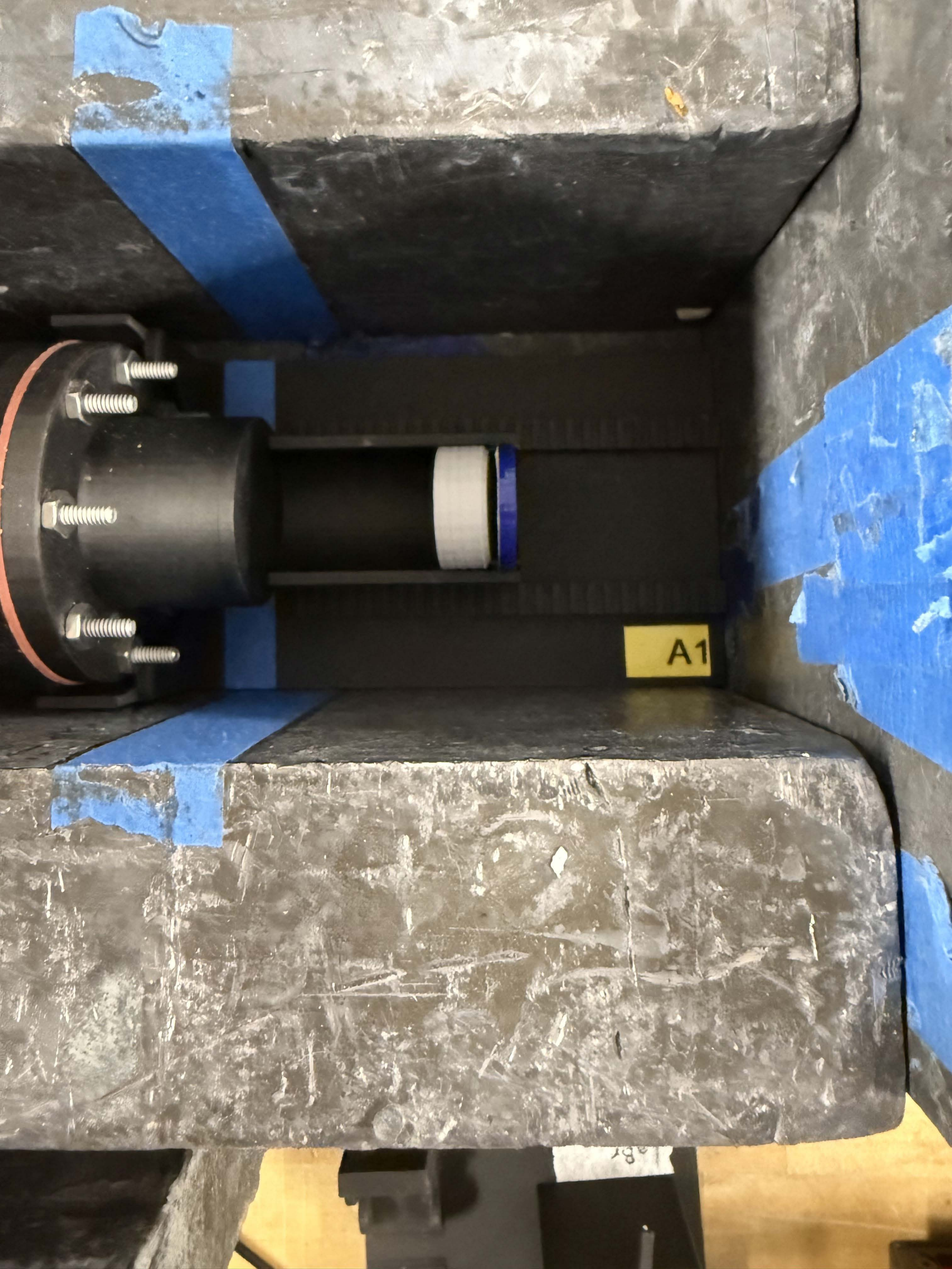}
\caption{Sample experimental setup for gamma source experiments}
\label{fig:setup:check}
\end{figure*}

The experimental setup is shown in~\cref{fig:setup:check}. A 3D-printed support aligns a \ce{^{137}Cs} gamma-ray source (\(E_{\gamma} = 662\)~keV) concentrically with the center of the \ce{LaBr_3} detector face and enables precise control over the source–detector distance. Additionally, a \ce{^{22}Na} (\(E_{\gamma}=511, 1275\)~keV) source is fixed on the support to verify that the position of the \ce{LaBr_3} detector remains unchanged across runs relative to the support.

Counts from the \ce{^{137}Cs} source were acquired under four configurations: unshielded and with each of the three capsule materials positioned between the source and detector. Each of these four measurements was repeated three times with the \gammaray source positioned in two different locations: 47.5 mm and 22 mm away from the detector. The percent attenuation for each capsule reported in~\cref{tab:gamma:attenuation} is computed as the average over the three data acquisitions. Each acquisition lasted for $5$ minutes, in which we acquired between $1.4\times10^5$ and $4\times10^5$ gamma counts under the \ce{^{137}Cs} peak. This corresponds to a relative error on the gamma counts less than $0.25$ \% due to the Poissonian statistics of the measurement. 

% Attenuation percentage table
\begin{table*}[b]
\centering
\captionof{table}{\label{tab:gamma:attenuation} Attenuation coefficients of $662$ keV $\gamma$-rays emitted by \ce{^{137}Cs} for four capsule materials. Coefficients are reported in \% with their absolute error.}
\smallskip
\begin{tabular}{l r}
\toprule
Capsule Material & Avg attenuation [\%]\\
\midrule
PETG	& $1.67\pm0.19$\\
PLA	    & $1.33\pm0.19$\\
PC	    & $1.32\pm0.19$\\
\bottomrule
\end{tabular}
\end{table*}

Attenuation coefficients for each capsule material are presented in~\cref{tab:gamma:attenuation}. All materials attenuate the 662 keV gamma rays by less than 1.7\%, and their values lie within two standard deviations of one another. While PETG exhibits marginally higher attenuation, the overall similarity suggests that the four materials would perform equivalently and introduce only negligible systematic error during foil counting on SPARC.

Finally, it is worth noting that the \emph{relative} error on each reported attenuation coefficient ranges between $10$ and $15$ \%, with some individual runs reaching up to $40$ \%. In preparation for SPARC operation, a more detailed characterization of the selected capsule material would be beneficial. This should include longer counting times, more measurement repetitions, and gamma attenuation across different energies by using other gamma-ray sources.

% Generator and Counting experimental setup
\begin{figure*}[!t]
\centering
\subfigure[]{\label{fig:experimental:setup:source}
\includegraphics[width=0.34\linewidth]{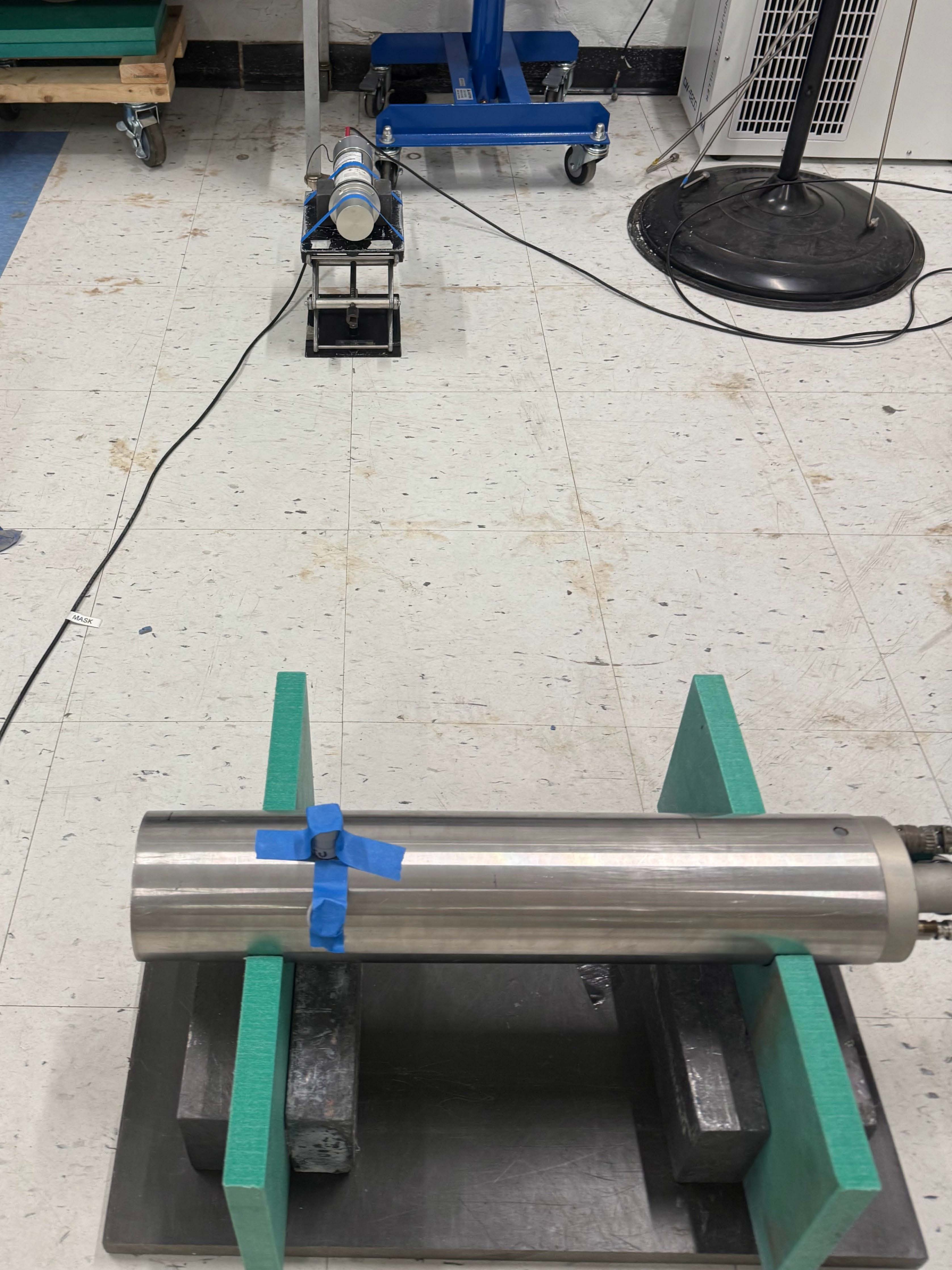}
}
\subfigure[]{\label{fig:experimental:setup:counting}
\includegraphics[width=0.6\linewidth]{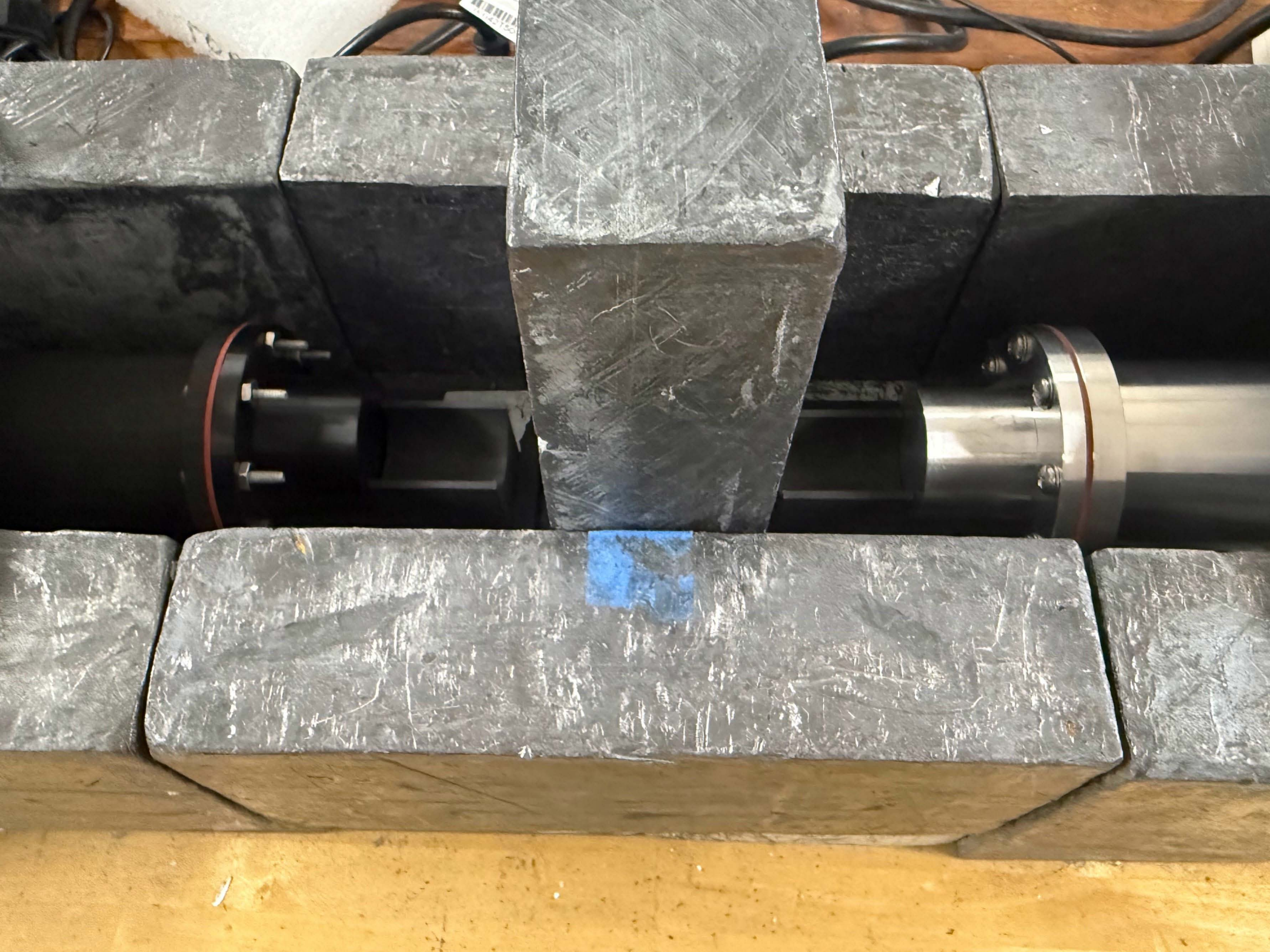}
}
\caption{
\subref{fig:experimental:setup:source} Experimental setup during foil irradiation. Foils are pushed against the bottom face inside the capsule (see fig.~\ref{fig:foil:holder}, which is then taped to the target plane of the DT neutron generator. Neutron rate is recorded by an EJ-301 liquid scintillator placed one meter away. 
\subref{fig:experimental:setup:counting} Experimental setup for measuring gamma spectra of irradiated foils. \(\text{LaBr}_{3}\) and \(\text{LaCl}_{3}\) detectors are used within a lead “sarcophagus” to minimize the background coming from the surrounding environment. Lead bricks also seal off detectors from the top (not pictured).}
\label{fig:experimental:setup:combined}
\end{figure*}

\section{Neutron irradiation}\label{sec:neutron:irradiation}
Foil irradiation experiments were conducted to quantify primarily the net attenuation (neutron attenuation during irradiation plus gamma attenuation during counting) introduced by different capsule materials, characterize the response capabilities of our detectors, and additionally verify the suitability of a multi-foil activation setup. Based on the selection criteria provided in~\cref{sec:foil:selection}, natural Al and Cu foils were selected for use in irradiation experiments. The foils' nuclear characteristics are detailed in~\cref{tab:foils:data:sheet} and their physical dimensions in~\cref{tab:foil:size}. A standardized procedure was followed for all experiments and is outlined below, along with descriptions of the materials and equipment used.

% Geometry of Foils
\begin{table*}[b]
\centering
\captionof{table}{\label{tab:foil:size} Geometric characteristics of a single foil of either Al or Cu (see fig.~\ref{fig:foils}). Thickness (T), Diameter (D), and Mass (M) are shown. Note, experiments were conducted with four foils stacked together to improve counting statistics.}
\smallskip
\begin{tabular}{l r r r}
\toprule
Material & T [$\mu$m] & D [mm] & M [mg]\\
\midrule
Al & 762 & 12.7 & 264\\
Cu & 127 & 12.7 & 140\\
\bottomrule
\end{tabular}
\end{table*}

\begin{figure*}[!t] 
\centering
\subfigure[]{
\label{fig:foil:holder}
\includegraphics[width=0.3\linewidth]{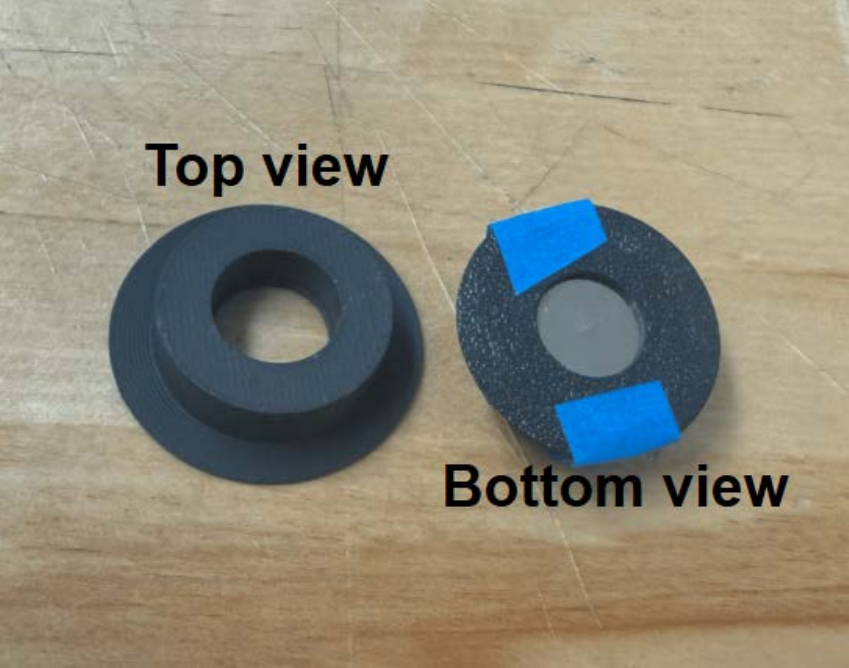}
}
\subfigure[]{\label{fig:3D:support:1}
\includegraphics[width=0.3\linewidth]{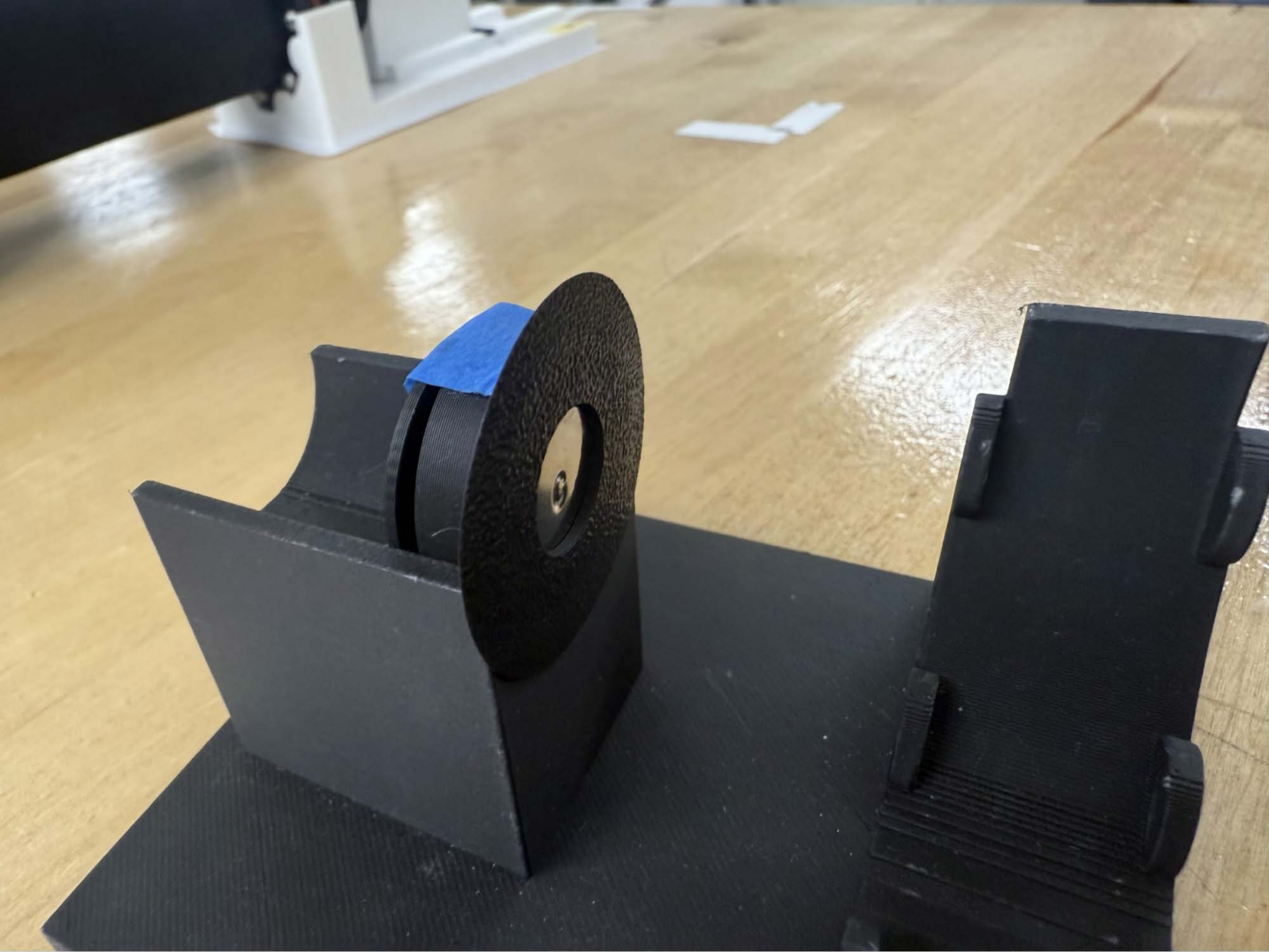}
}\subfigure[]{\label{fig:3D:support:2}
\includegraphics[width=0.3\linewidth]{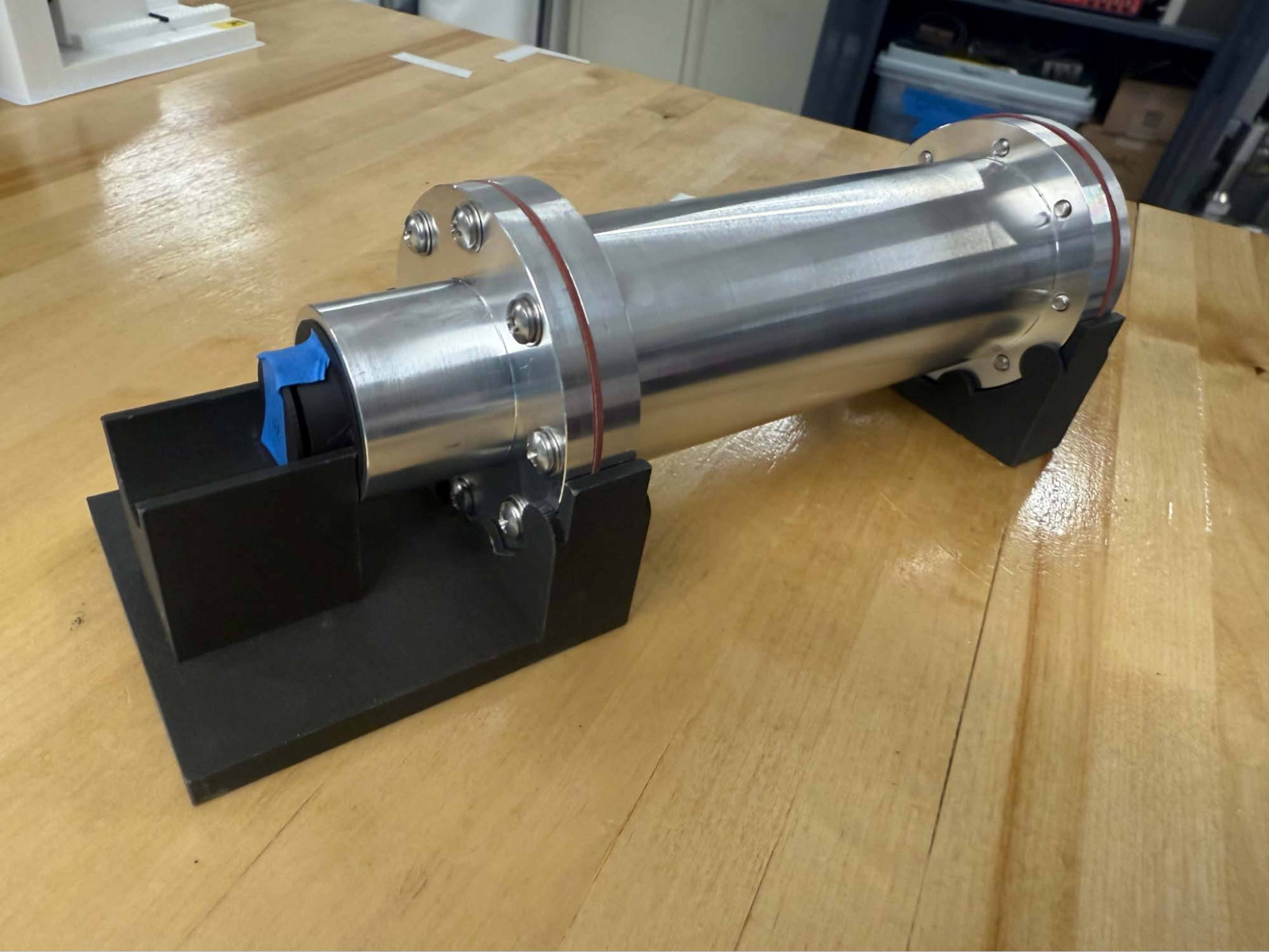}
}\\
\caption{\subref{fig:foil:holder} Foil holder ensures the same distance from the detector for capsule and no capsule experiments. 
\subref{fig:3D:support:1} Foils inside a printed foil holder, mounted on detector support. \subref{fig:3D:support:2} Example geometric setup of counting a no-capsule experiment; foil holder is replaced by capsule for experiments involving capsules, both push against the face of the detector}
\label{fig:3D:support}
\end{figure*}

Four foils are stacked together in either a standard capsule or a custom-designed holder that replicates the capsule cap's thickness (0.05 inches) without introducing additional material interference (see~\cref{fig:foil:holder}). In both scenarios, the holder or capsule is taped to the surface of a DT generator along the target plane, as seen in fig.~\ref{fig:experimental:setup:source}. In order to be efficient with irradiation and counting time, we focus our analysis on the reactions with shorter half-lives: \(^{27}\text{Al}(n,p)^{27}\text{Mg}\) and \(^{63}\text{Cu}(n,2n)^{62}\text{Cu}\) (\cref{tab:foils:data:sheet}). These reactions approach 98.7\% of their saturation activity after a 1-hour irradiation, which is why we adopted a 1-hour duration for each irradiation and counting.
During irradiation, an EJ-301 liquid scintillator, positioned one meter from the DT source, monitors the stability of the neutron generator (see~\cref{fig:experimental:setup:source}). Once an irradiation is complete, the foils are promptly transferred to a separate room for counting, which begins 120 seconds after the end of irradiation. Al foils are counted with \(\text{LaBr}_{3}\) and Cu foils with \(\text{LaCl}_{3}\) each for 1-hour (see~\cref{fig:experimental:setup:counting}); detector properties are detailed in~\cref{tab:detector}.  3D-printed supports hold the foil-holder or capsule concentric and in contact with the detector face. This ensures consistent geometric conditions across all experiments (see~\cref{fig:3D:support}).

% 3D printed supports

\begin{figure*}[!b]
    \centering
    \includegraphics[width=0.5\linewidth]{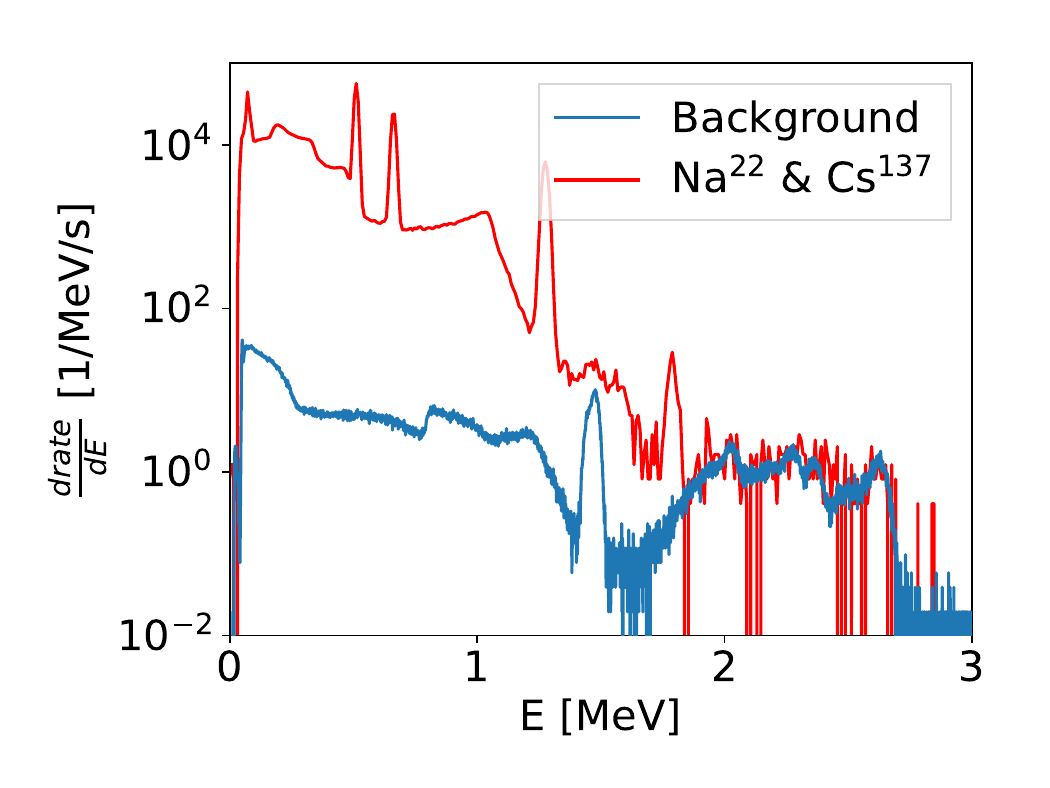}
    \caption{\gammaray spectrum of \ce{^{22}Na} and \ce{^{137}Cs} sources measured with the \LaBr detector, compared with the spectrum of the environmental background measured with the same detector.}
    \label{fig:calibration:spectrum}
\end{figure*}

% Detector parameters
\begin{table*}[b]
\centering
\captionof{table}{\label{tab:detector} \ce{LaBr_3} and La\ce{Cl_3} detector properties. Diameter (D) and height (H) given in cm; photomultiplier tube voltage (\ce{V_{PMT}}) varied by experiment; detector resolution (R) at 662 keV obtained from a \ce{^{137}Cs} gamma source.}
\smallskip
\begin{tabular}{l r r r r r r}
\toprule
Detector & Foil & D [cm] & H [cm] & \ce{V_{PMT}} [V] & R @ 662 keV \\
\midrule
La\ce{Br_3} & Al & 2.5 & 2.5 & 600–1100 & 3.2\%\\
La\ce{Cl_3} & Cu & 2.5 & 2.5 & 700–1100 & 4.5\%\\
\bottomrule
\end{tabular}
\end{table*}

 A consistent DT generator source voltage of -90 kV and +3 kV source acceleration [tbd] is used across experiments. The DT generator has a nominal maximum neutron rate production of \( \approx 10^8\,\text{n/s} \). The energy spectrum of the source is strongly influenced by the neutron emission angle~\cite{ball2024} due to Doppler shift. At $180$ deg with respect to the D beam direction, the average emitted energy was measured to be $13.65$ MeV. In this study, we placed both the foils and the neutron monitor alligned with the target plane of the generator. In that position they intercept the emission at an angle of $90$ deg, at which the average energy of the source was measured to be $14.1$ MeV. 
 
 All three detectors were calibrated with \ce{^{22}Na} and \ce{^{137}Cs}  gamma sources before each experiment as small changes in environmental conditions can affect detector performance. A 14 hour background measurement was also performed following each counting. The lead shielding arranged around the \gammaray detectors, reduced the background rate from \( \sim150~\text{counts/s} \) to \( \sim30~\text{counts/s} \). The spectra of both the calibration sources and the environmental backgrond are shown in~\cref{fig:calibration:spectrum}.

\begin{figure*}[!t] 
\centering
\subfigure[]{\label{fig:calibration:energy}
\includegraphics[width=0.5\linewidth]{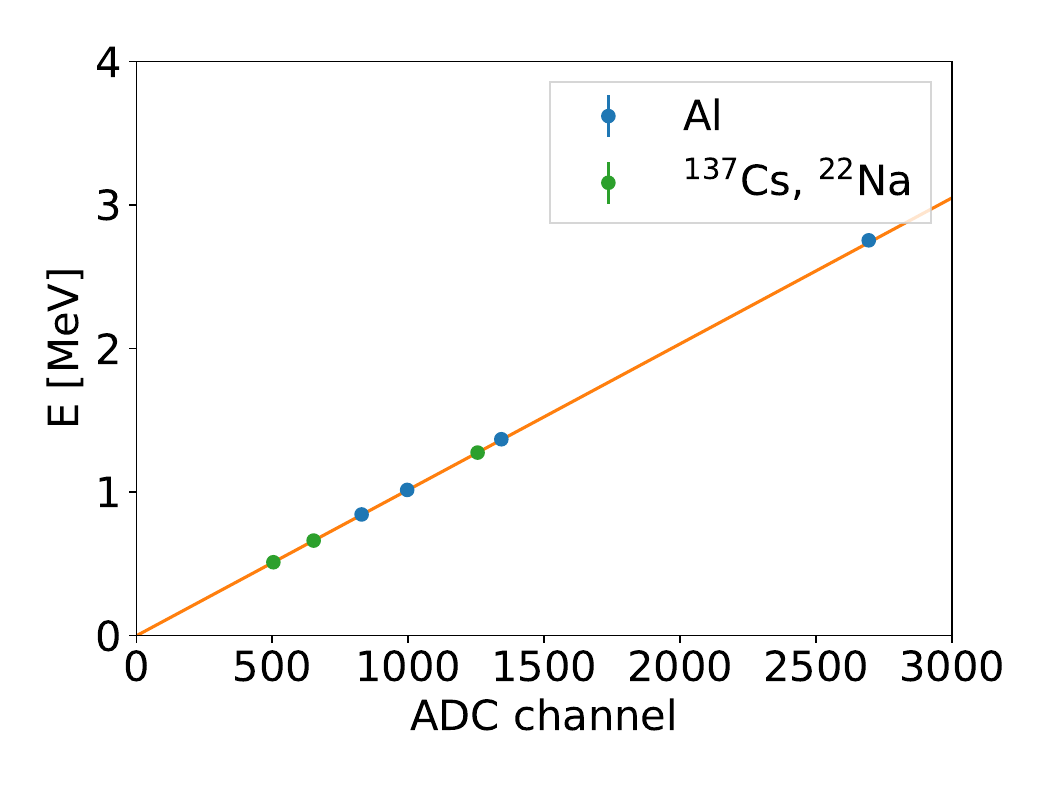}
}\subfigure[]{\label{fig:calibration:resolution}
\includegraphics[width=0.5\linewidth]{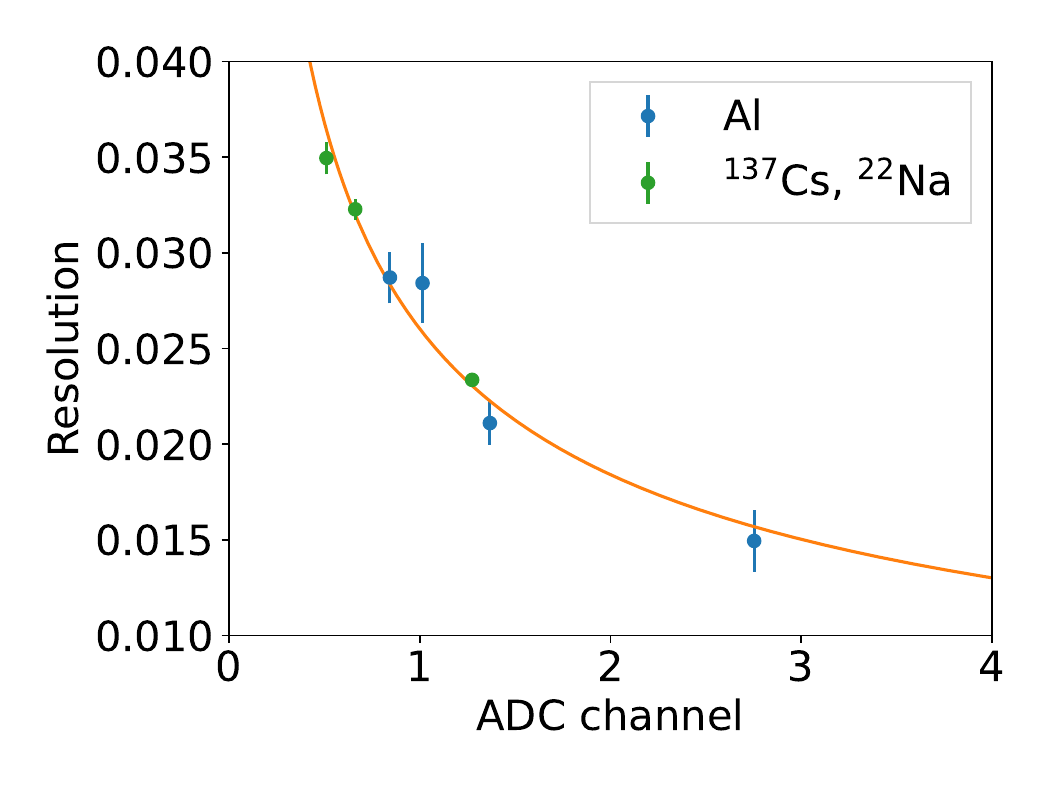}
}\\
\caption{\subref{fig:calibration:energy} Linear calibration of the \LaBr detector at a 700 V PMT voltage. Calibration performed with the three gamma peaks emitted by \ce{^{137}Cs} and \ce{^{22}Na}
 gamma-ray sources. Peaks of Al (844, 1014, 1369, and 2754 keV) are overlaid on the fit (orange line) to verify the calibration.
\subref{fig:calibration:resolution} Energy resolution of the \LaBr detector. The experimental resolution, calculated from the gaussian fit of the Al gamma-ray peaks, is shown in blue round dots. The experimentally-found resolution is fit to formula $k / E$ with $k$ free parameter. The result is shown as an orange solid line.}
\label{fig:calibration}
\end{figure*}

A DT5730s (500 MSps, 14 bit) digitizer from CAEN~\cite{CAENdigitizer} was used along with CoMPASS software ~\cite{CAENcompass} to aquire data. Post-analysis was conducted with an in-house Python library developed for detector raw data analysis, called Daisy~\cite{panontin2024}. Daisy is used to calibrate energy spectra (see~\cref{fig:calibration:energy}), determine energy resolution (see~\cref{fig:calibration:resolution}), and fit peaks while subtracting background (see~\cref{fig:foil:spectrum}). To retrieve the counts under a peak, a region of interest is first manually selected. Daisy then fits the peak with a Gaussian function, integrates it analytically, and estimates the background by referencing the count value on either side of the peak. The uncertainties introduced from manual selection of the region of interest is discussed in~\cref{app:error:normalized:counts}.

% Calibration

% Calibration spectrum
%\begin{figure}[H] 
%\centering
%\includegraphics[scale=0.5]{figures/Cu_in_PC_Capsule_zoom.png}
%\caption{Spectrum of gamma-rays emitted by Cu foils after 1 hr long irradiation inside a PC capsule. The Gaussian fit of the 511 keV peak and the linear fit of the background to the peak are also shown.}
%\label{fig:foil:Fitting}
%\end{figure}

% Experiments performed
\begin{table*}[!t]
\centering
\captionof{table}{\label{tab:experiments} List of neutron irradiation experiments performed.}
\smallskip
\begin{tabular}{l l r l}
\toprule
 & Run ID & Irradiation time & Description\\
\midrule
No Capsule & 1 hr & 1 hr & Standard irradiation and counting (Al-La\ce{Br_3}, Cu-La\ce{Cl_3})\\
 & 1.5 hr & 1.5 hr & Longer irradiation\\
 & Multi-foil & 1 hr & 4 Al and 4 Cu foils stacked in foil holder together\\
 & Swap & 1 hr & Al counted by La\ce{Cl_3} and Cu counted with La\ce{Br_3}\\
\midrule
With Capsule & PETG & 1 hr & Standard irradiation and counting\\
 & PLA & 1 hr & Standard irradiation and counting\\
 & PC & 1.5 hr & Longer irradiation\\
\bottomrule
\end{tabular}
\end{table*}

The complete list of irradiation experiments performed is outlined in~\cref{tab:experiments}.
In all but one experiment, Al was counted with LaBr\textsubscript{3} and Cu with LaCl\textsubscript{3}. To test the multi-foil configuration, Al and Cu foils were irradiated and counted together within the same foil holder. Each foil holder contained a total of eight foils: four Al and four Cu foils stacked in alternating layers (Al–Cu–Al–Cu, etc.). Two identical holders were employed, each assigned to a dedicated detector for independent counting. Moreover, to enable direct comparison between the multi-foil and single-foil experiments, an additional experiment was performed in which the detectors were "swapped": Cu was counted with LaBr\textsubscript{3} and Al with LaCl\textsubscript{3}. For capsule experiments, PETG, PLA, and PC capsules were tested. Finally, a longer 1.5-hour irradiation was conducted to assess whether secular equilibrium was reached for the short-lived activation products. 

%%%%%%%%%% NEW SECTION %%%%%%%%%%%%
\subsection{Neutron Irradiation Results \& Discussion}\label{subsec:Neutron:Results}

In this section, we present and discuss the results of gamma-ray spectral analysis and the stability of the neutron generator. We then provide the corresponding foil gamma-count data and discuss its implications for SPARC.

\begin{figure*}[!b] 
\centering
\subfigure[]{\label{fig:foil:spectrum:al}
\includegraphics[width=0.5\linewidth]{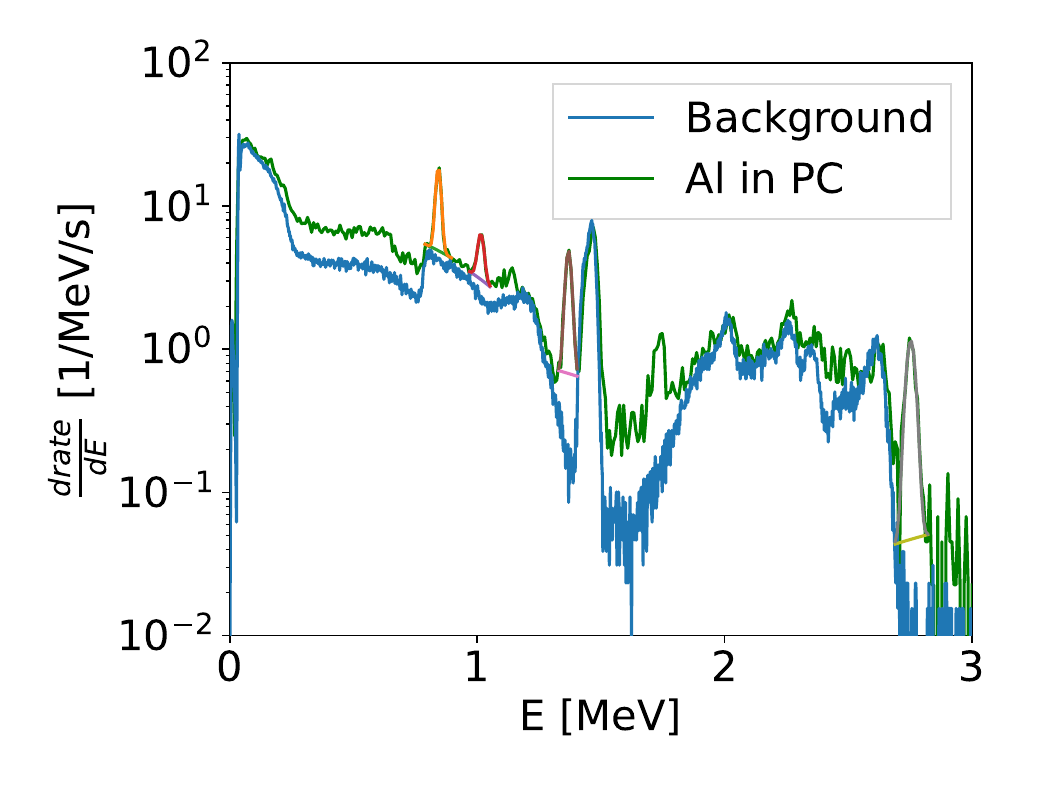}
}\subfigure[]{\label{fig:foil:spectrum:cu}
\includegraphics[width=0.5\linewidth]{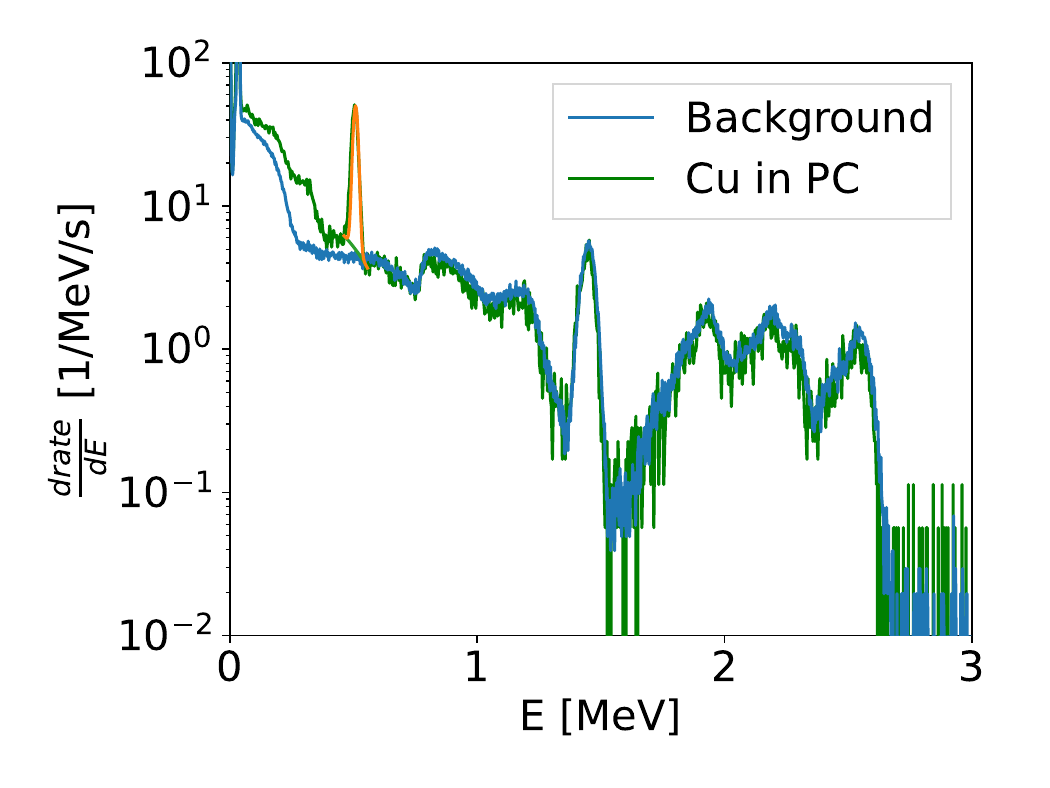}
}\\
\caption{\subref{fig:foil:spectrum:al} Gamma spectrum of four irradiated foils in a polycarbonate rabbit. a) is Al counted by a \ce{LaBr_3} scintillator and b) is Cu counted by a \ce{LaCl_3} scintillator. The gaussian fit, with a linear background estimate, is shown over the peaks.}
\label{fig:foil:spectrum}
\end{figure*}

\begin{figure}[!t]
\centering
\subfigure[]{\label{fig:foil:multi:lacl}
\includegraphics[width=0.48\linewidth]{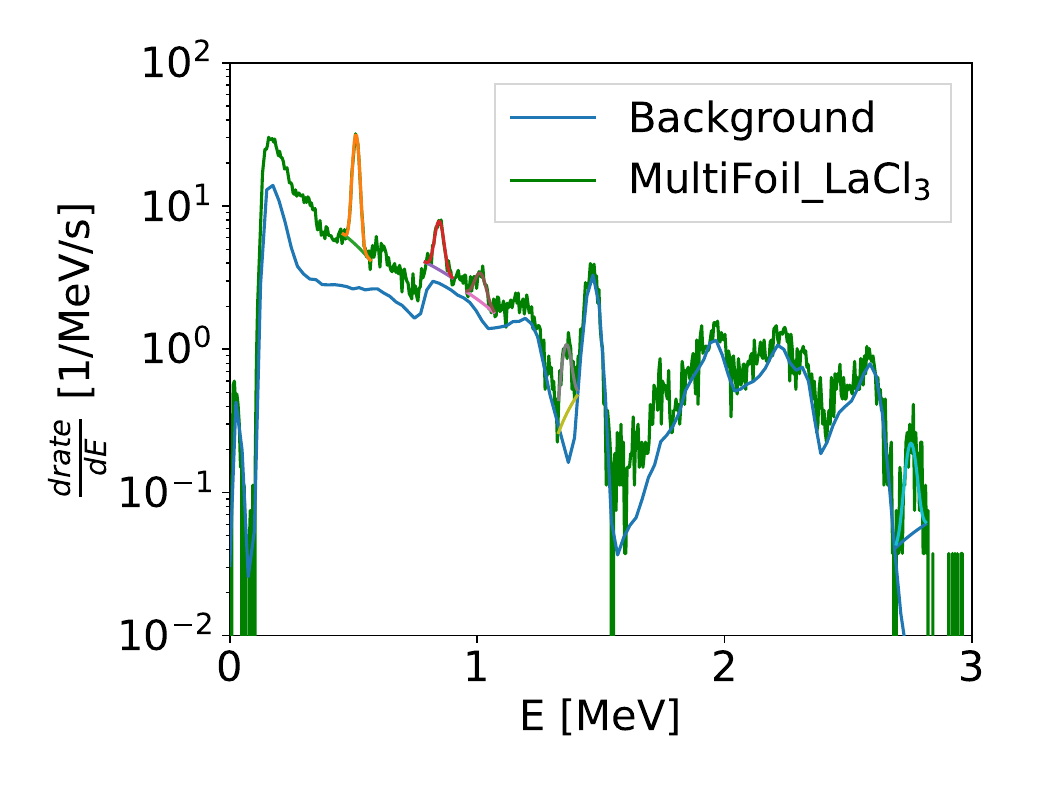}}
\subfigure[]{\label{fig:foil:multi:labr}
\includegraphics[width=0.48\linewidth]{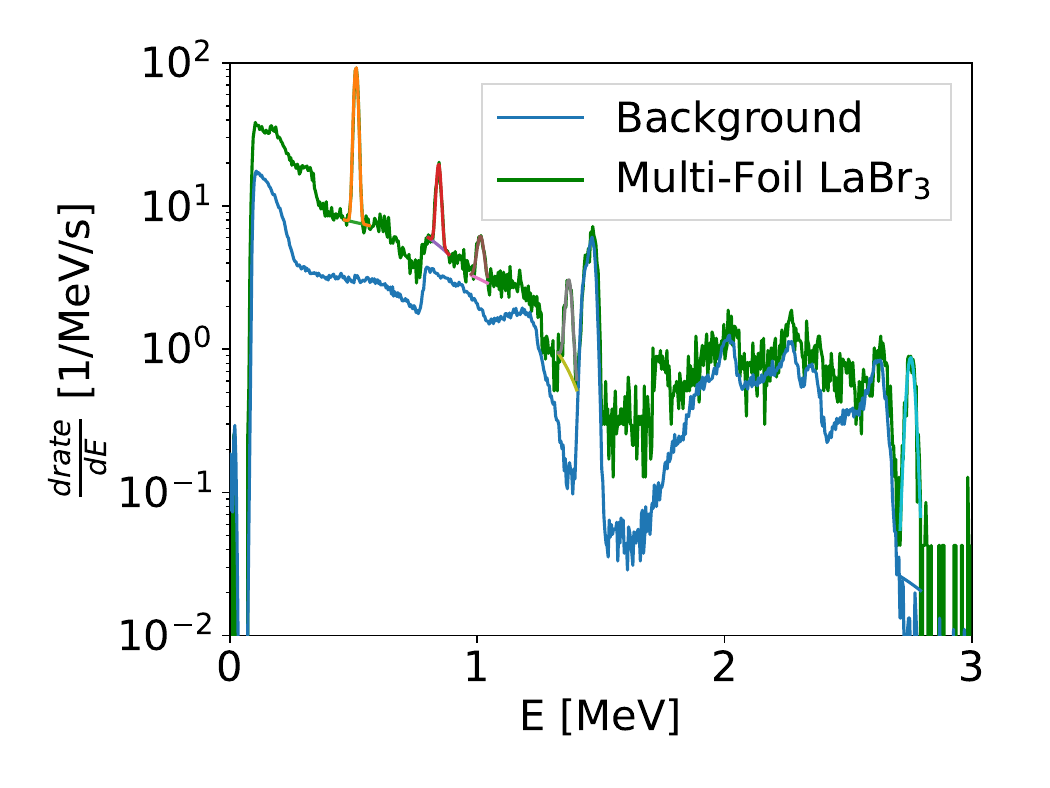}}
\caption{Gamma spectra of Al and Cu foils in 'Multifoil' configuration after a 1~hr irradiation, measured by \subref{fig:foil:multi:lacl} a \ce{LaCl_3} scintillator and \subref{fig:foil:multi:labr} a \ce{LaBr_3} scintillator.}
\label{fig:foil:multi}
\end{figure}

\Cref{fig:foil:spectrum} displays two representative gamma-ray spectra for foil irradiation experiments. Al exhibits four characteristic peaks (844, 1015, 1369, 2754 keV), while Cu shows one (511 keV). Both \LaBr and \LaCl detectors can resolve all $5$ peaks emitted by the two activated foil materials. The higher efficiency of \ce{LaBr_3} is clearly an advantage as it increases the statistics of the foil counting. At the same time \LaBr has a higher capacity to resolve peaks with similar energies, as discussed in~\cref{sec:gamma:attenuation}. Both detectors exhibit an intrinsic background with a prominent peak at $\approx1465$ keV, which in our experiments has roughly the same count rate as the 1369 keV peak from Al. The peak is emitted by \ce{^{168}La} contained in the scintillation crystal, so it can limit the capacity of these detector to measure foils with \gammaray peaks at similar energies. In our experiments, we noticed that the higher resolution and efficiency of \LaBr helped the separation of the Al-induced 1369 keV peak from the background-induced $1465$ keV peak in~\cref{fig:foil:multi}. In terms of suitability for SPARC, an HPGe detector remains the preferred option for multifoil activation on SPARC given its improved resolution, although we have shown that \LaBr is also capable of performing adequately.

% Foil spectra

The validity of a linear energy calibration based on \ce{^{137}Cs} and \ce{^{22}Na} gamma-ray sources was tested. \Cref{fig:calibration:energy} shows the four Al peaks overlaid on the linear fit derived from the gamma-ray source energies. The results indicate at most a 2\% deviation between the measured peak energies and their expected values. This is valid also for the 2754 keV peak peak from Al, proving the validity of the extrapolation at high energies of the linear energy calibration. 

The experimental peaks are fit with a gaussian peak on top of a linear background to extract their energy and the measured counts under the peak. This fitting strategy works well for the lower energy peaks. 
In~\cref{fig:foil:spectrum:al}, we clearly see that the background of the 1368 keV and 2754 keV peaks has a non trivial functional form. 
We are currently developing a more precise fitting algorithm based on a synthetic diagnostics approach capable of fitting the whole measured spectrum. 
For the sake of the present work, however, we note that the gaussian + line fit converges also for the 1368 keV and 2754 keV peaks and can give a first estimate of their counts. 
The systematic error related to this fitting strategy has been assessed by varying the bin width and the peak fitting region of interest in post-analysis. 
The reported counts for the 1369 keV and 2754 keV peaks can fluctuate by over 20\%, whereas the 844 and 1015 keV peaks always remained within a stable 5\% range. These discrepancies were observed across multiple experiments, supporting our decision to exclude the two longer-lived reactions from the Al gamma-count analysis.

\begin{figure*}[!b] 
\centering
\includegraphics[width=0.7\linewidth]{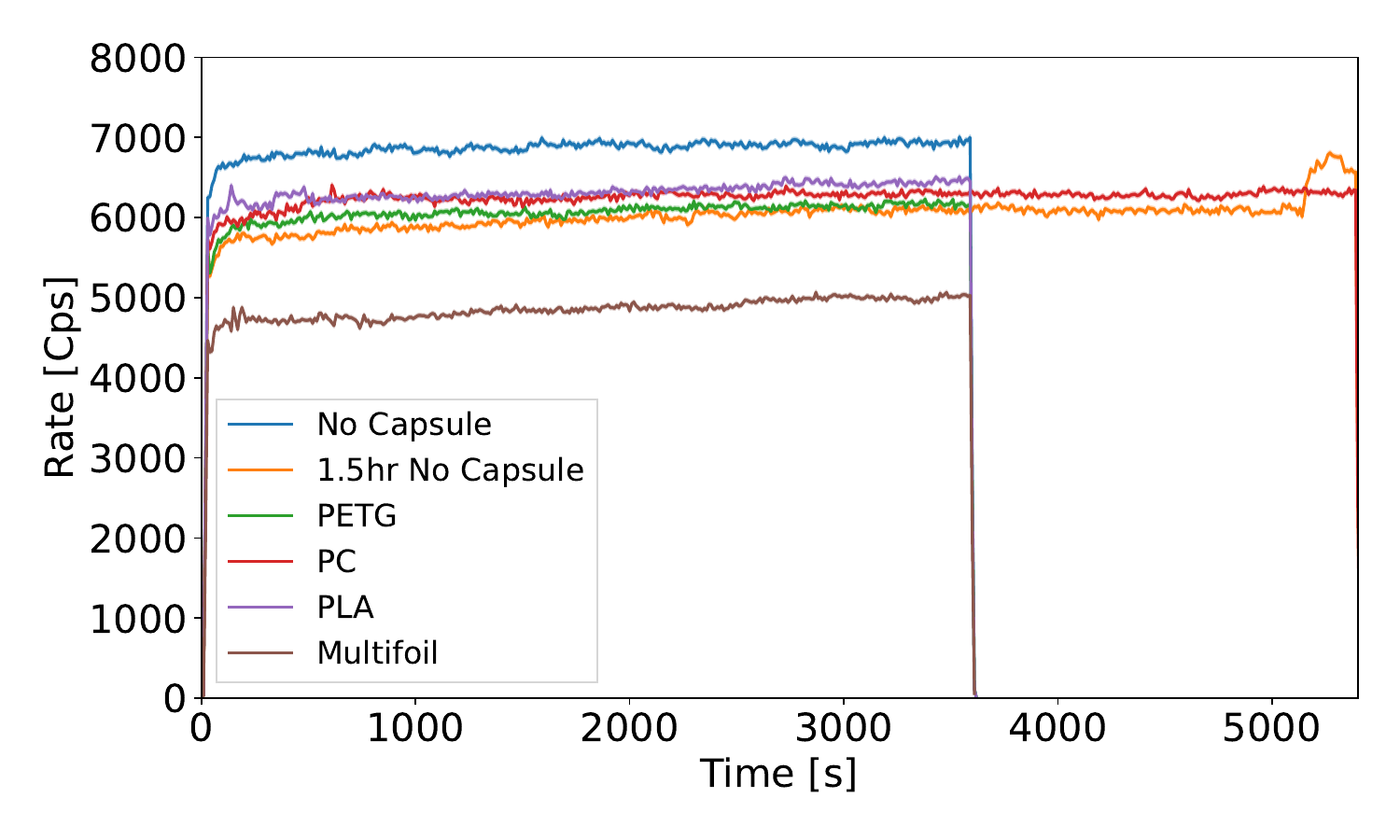}
\caption{Neutron rate for select experiments, as measured by a deutered liquid scintillator positioned 1 meter away from the DT generator (see fig.~\ref{fig:experimental:setup:source})
%PC duration extends to 5400 seconds but not show
}
\label{fig:dt:rate}
\end{figure*}

Because the nuclear reactions analyzed in this study approximately reach secular equilibrium during irradiation, gamma activity scales linearly with neutron rate, see~\cref{eq:foil:irradiation:2}. 
Therefore, gamma counts per neutron rate is chosen as the most reliable basis for comparison between experiments. 
\Cref{fig:dt:rate} shows the neutron rate during irradiation for select experiments.
The measured neutron rate from experiment to experiment can span from about $5\times10^3$ to $7\times10^3$ Cps, see for instance the multifoil and the No Capsule time traces.
This behavior is observed even if the neutron generator was operated at constant source and accelerator voltages.
In some experiments, we also noticed occasional instability of the generator operation, which showed sudden \textasciitilde10\% changes in neutron rate (see for instance the 1.5 hr No Capsule time trace). 

\begin{figure*}[!b] 
\centering
\includegraphics[width=0.6\linewidth]{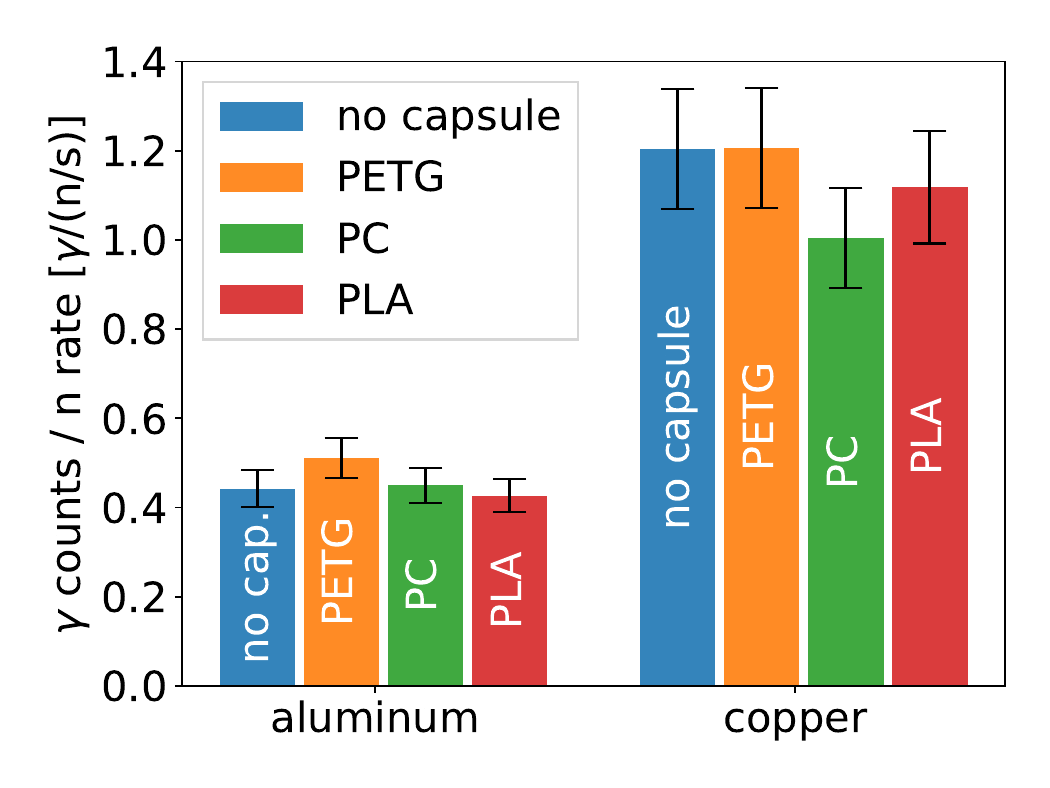}
\caption{Results of foil irradiation experiments performed. ’Al counts’ is the sum of counts from 662 keV and 1015 keV peaks, divided by neutron rate. ’Cu counts’ is solely counts of 511 keV peak, divided by neutron rate.}
\label{fig:results:barchart}
\end{figure*}

These variation in the neutron rates are ascribable to oscillations in the performance of the DT generator. 
They can also be due to changes in the materials from the other experiments in the shared lab space, which in turn can change the fraction of scattered events measured.
We also notice a slight increase of neutron rate throughout the irradiation period is likely due to activation of the surrounding material.
As shown in~\cref{eq:foil:irradiation} and~\cref{eq:foil:counting}, a change in neutron production will necessarily affect also the final number of gamma-rays counted with the inorganic scintillators. In order to compare the results of different irradiations, we can use~\cref{eq:foil:irradiation:2} and normalize the total gamma counts by neutron rate measured by the neutron monitor. All collected data are compiled in~\cref{tab:results} and~\cref{fig:results:barchart}.

%Gamma counts, on the other hand, do not include any geometric uncertainty as the 3D-printed supports minimized this. Gamma-ray measurements are subject to their own statistical uncertainty ($\sqrt{C}$) at each energy peak, with manual peak fitting routines introducing an additional variability of approximately 5\%. Furthermore, variations in foil mass and rabbit print quality are considered insignificant due to a tight $\pm0.0001$ g tolerances on foils from manufacturer [shieldwerks] and visually identical rabbits. 
%Uncertainty total 0.32 ± 0.03= (need to re derive)

% DT rates

In order to compare adequately \gammaray counts / neutron rate across experiments, we evaluate each measurement's distinct statistic and systematic sources of error. Ultimately we combine them to produce a net uncertainty per experiment. The detailed derivation of these errors is reported in~\cref{app:error:normalized:counts}. 
The final results of this analysis are reported in~\cref{tab:results} and~\cref{fig:results:barchart}. 
The normalized number of counts of irradiations performed with and without plastic capsules are within 1 standard deviation from each other.
This result is in line with the low gamma attenuation coefficients measured in~\cref{sec:gamma:attenuation}.
It also shows that the neutron attenuation due to the capsules is significantly lower than the systematics uncertainties introduced by the present experimental setup.

% Table of RESULTS
\begin{table*}[!t]
\centering
\captionof{table}{\label{tab:results} Results of foil irradiation experiments performed at the MIT Vault Lab. 'Al counts' is the sum of counts from 662 keV and 1015 keV peaks summed together [SHOULD BE, SUM OF ALL FOR NOW. CHANGE]. 'Cu counts' is solely counts of 511 keV peak. 'n rate' is average neutron rate during irradiation}
\smallskip
\begin{tabular}{l r r r}
\toprule
Run ID & Al [counts / (n/s)] & Cu [counts / (n/s)] & n rate [n/s]\\
\midrule
1 hr    & $0.44 \pm 0.04$ & $1.20 \pm 0.13$ & $6833 \pm 137$ \\
1.5 hr  & $0.52 \pm 0.05$ & $1.34 \pm 0.15$ & $6000 \pm 120$ \\
Swap    & $0.32\pm0.03$ & $1.93 \pm 0.21$ & $6250 \pm 125$ \\
Multi-foil \LaBr & $0.53 \pm 0.06$ & $2.05 \pm 0.25$ & $4833 \pm 97$ \\
Multi-foil \LaCl & $0.28 \pm 0.04$ & $0.90 \pm 0.10$ & $4833 \pm 97$ \\
PETG    & $0.51 \pm 0.05$ & $1.21 \pm 0.13$ & $6028 \pm 121$ \\
PLA     & $0.43 \pm 0.04$ & $1.12 \pm 0.13$ & $6228 \pm 126$ \\
PC      & $0.45 \pm 0.04$ & $1.00 \pm 0.11$ & $6222 \pm 124$ \\
\bottomrule
\end{tabular}
\end{table*}

% Results bar plot

It is worth discussing in depth the comparison between the "Multi-foil", the "1 hr" and the "Swap" runs.
The gamma-ray counts under the Cu and Al peaks during multifoil experiments match those obtained from individual Al-only and Cu-only experiments ("1 hr" and "Swap") within error. 
This validates the use of Al+Cu multifoils measurements for SPARC, which could be used to validate the reconstruction of the total fusion energy produced by SPARC by comparing the analysis of two different materials with different activation paths. 
We also note that \LaBr measurements have a higher statistics than \LaCl ones. This is true when comparing the "Swap" run with the "1 hr" run and when comparing the two "Multi-foil" measurements.
The "Multi-foil" experiment was performed by irradiating one set of foils and placing the two detectors back to back, with the Al and Cu foils in between them. 
The \LaBr detector has custom made plastic casing, while the \LaCl detector has a aluminum casing.
The foil \gammaray emission can be considered isotropic and the solid angles subtended by the two detectors in the three runs is comparable. 
The difference in the statistics of the \LaBr and \LaCl measurements, then, is mostly due to the different efficiencies of the two detectors as well as to the different casing materials. 
A detailed analysis of the two contributions is beyond the scope of the present paper, however this comparison further shows how \LaBr can outpeform \LaCl detectors for foil counting.

Finally, the comparison between the "1 hr" and the "Swap" runs shed some light on the performance of Al and Cu foils. In the same experimental conditions, Cu foils yield at least 4 times more photons than Al foils (see~\cref{tab:results}). If we normalize the number of counts per total mass of the foils used in our experiments (see~\cref{tab:foil:size}), Cu foils yields about 8 times more \gammaray counts than Al foils. At the same time, Cu emitts $511$ keV \gammaray via two possible activation paths (see~\cref{tab:foils:data:sheet}), which can complicate the post-analysis of the experimental data on SPARC and the reconstruction of the total fusion energy produced. For single foil measurements, we can then imagine to use Al as the baseline foil for $14.1$ MeV neutron detection on SPARC. On the other hand, Cu foils can be useful in low power experiments to increase the statistics of the FOIL measurement by almost an order of magnitude.

\section{Conclusions}\label{sec:conclusions}
This work characterizes the various components of a multi-foil activation system. The system was tested on a DT neutron generator in order to inform the design of a dedicated activation foil diagnostics for the SPARC tokamak. We performed FISPACT simulations of realistic irradiation conditions on SPARC and at the MIT laboratories for various materials. The results suggested to focus our experimental activities on Al and Cu foils, which can be used on SPARC to detect $14.1$ MeV neutrons from DT fusion reactions and have an expected \gammaray yield high enough to be measurable in the MIT laboratories. Based on these simulations, possible future experimental campaigns could consider materials such as Fe, to test multi-foil experiments aimed at assessing the energy distribution of the neutron spectrum between $5$ and $14$ MeV.

We then established a complete experimental testing procedure for foil irradiation and counting. The experiments were conducted at the MIT laboratories, using a DT neutron source to irradiate Al and Cu foils. We further deployed a EJ301 liquid scintillator to monitor the stability of the neutron generation rate and correct for any variation between runs. We used \LaBr and \LaCl spectrometers to count the \gammaray spectrum emitted by activated foils. A detailed analysis of both statistic and systematic error sources is presented for both the irradiation and the counting phase. For all experiments, the relative error on the total measured \gammaray counts normalized per neutron rate is higher than $10$ \%. This error levels would be too high for SPARC, where we aim at reconstructing the total $E_{fus}$ produced with a total uncertainty of $10$ \%, combining together errors in the calibration of the FOIL system, in the neutronics simulations and in the experimental measurements. The uncertainty calculations presented in the appendix can be used to dimension the FOIL system (e.g. foil mass, position, transportation time, counting time, detector type, tolerances, etc) to lower the experimental uncertainty to an acceptable level. 

To support the design an automated pneumatic transport system to move the activated foils between the SPARC torus and the \gammaray detectors, we tested 3D-printed thermoplastic capsules . We considered three different materials to assess their effects on a foil counting experiment. In particular we used a \ce{^{137}Cs} \gammaray source to assess the \gammaray attenuation coefficient, which was measured to be below 2\% for all three materials. We also performed neutron irradiations with Al and Cu foils contained inside the capsules, showing that the attenuation of the neutron spectrum is well below the experimental uncertainties on the final number of \gammaray counts measured. We can then conclude that the capsules have a negligible effects on the experiments we conducted. Further Monte Carlo simulations will be needed to assess the absolute attenuation coefficient of these capsules and compare it with the other expected systematic uncertainties on SPARC experiments.

We have also studied experimentally a multi-foil configuration with Cu and Al irradiated and measured together. 
As expected, this test yielded consistent results with single-foil experiments, where Al and Cu were irradiated and measured independently. 
In particular \LaBr and \LaCl were capable of resolving the 6 peaks contained in the combined spectrum of Al and Cu foils (4 peaks from Al, 1 peak from Cu and 1 peak from \ce{^{168}La} intrinsic background). 
However, it is worth noting that the higher efficiency of \LaBr, granted higher statistics under each peak and facilitated the convergence of the fitting procedure. 
This suggests the necessity of studying the stability of the fitting procedure for a realistic SPARC case, possibly using a synthetic diagnostics tool validated on current experiments.

Finally, we have investigated the possibility of using alternative detectors to the traditional HPGe for measuring the \gammaray decay spectrum of the irradiated foils. 
For all the experiments here reported, we used 1 x 1 \ce{inch^2} (diameter x height) \LaBr and \LaCl inorganic scintillators. 
The advantages of inorganic scintillators are a lower price with respect to HPGe and the capacity of operating at room temperature. 
Their drawback is a ten times lower energy resolution. 
Based on the three detector resolutions, we assess their capacity of distinguish close \gammaray peaks in multi-foil experiments. 
Even though they have a lower resolution, \LaBr and \LaCl can resolve a good number of foil-mixtures. 
In our experimental tests \LaBr emerged as the best detector for foil counting, thanks to its higher efficiency with respect to \LaCl. In particular, \LaBr outperforms \LaCl in measuring Al, Cu, and multi-foil spectra, proving to be a compelling alternative to HPGe.

\subsection*{Acknowledgments}
This work was supported by Commonwealth Fusion Systems and by the FUSAR program organized by MIT. We extend our special thanks to M. Short and R. Shulman for their leadership of the FUSAR program at MIT. We are also grateful to T. Hagenlocker and E. Hopkins for their collaboration. We thank Z. Hartwig for granting us access to the Vault laboratory and DT generator, and finally J.L. Ball, S. Mackie, L. Nichols and D. Pettinari for their valuable support. 

\appendix
    \section{The SPARC tokamak}\label{app:sparc}

SPARC is a high-field tokamak currently under construction by Commonwealth Fusion Systems (CFS) that aims to demonstrate a fusion gain factor \( Q_{\text{fus}} > 1 \). During the flat-top phase of the Primary Reference Discharge (PRD)~\cite{creely2020}, SPARC is expected to reach a neutron yield rate \( \frac{dY_n}{dt} \) up to \( 5 \times 10^{19} \)~n/s \cite{raj2024}, corresponding to a DT fusion power of 140~MW. The majority of neutrons generated in the plasma will have an energy of 14.1~MeV from deuterium-tritium (D-T) reactions and a much smaller fraction of neutrons an energy of 2.5~MeV from deuterium-deuterium (D-D) reactions. 

SPARC is planning four main neutron diagnostic systems~\cite{raj2024}.
A neutron camera~\cite{ball2024} and a proton recoil spectrometer~\cite{mackie2024, dallarosa2024} will be placed behind collimated lines of sight outside of the tokamak hall. 
Together, these systems will perform neutron spectroscopy and infer the neutron emissivity profile, total fusion power (\(P_{\text{fus}}\)), and information about the ion populations, such as temperature, fuel ratio, and the distribution of supra-thermal ions. 
Neutron monitors~\cite{wang2024} will be placed inside the tokamak hall to measure the neutron flux \( \varphi_{\text{n}} \) and total fusion rate.
SPARC will also deploy activation foils, which will be irradiated inside a port plug, providing a pulse-by-pulse measurement of the neutron fluence and total fusion energy. 
    \section{Details on purchased foils}\label{app:foil:purchase}

    The Cu and Al foils were purchased from the company Shieldwerx. Details on the purity and composition are provided in~\cref{tab:Cu:comp,tab:Al:comp}.

    \begin{table}[!h]
        \centering
        \caption{Cu foil samples.}
        \label{tab:Cu:comp}
        \begin{tabular}{c r}
            Element & (\%) \\
            \hline %\\
            Cu & 99.9928 \\
            Fe & 0.0010 \\
            Ni & $<0.0050$ \\
            Cr & 0.0001 \\
            Al & $<0.0010$ \\
            Pb & 0.0004 \\
            Sn & 0.0004 \\
            Zn & $<0.0010$ \\
            Si & 0.0010 \\
            Ag & 0.0010 \\
            Mg & 0.0002 \\
            Ca & 0.0002 \\
            Bi & 0.0004 \\
            \hline
        \end{tabular}
    \end{table}

    \begin{table}[h!]
        \centering
        \caption{Al foil samples.}
        \label{tab:Al:comp}
        \begin{tabular}{c r}
             Element & (\%) \\
            \hline %\\
            Al & 99.9901 \\
            Cu & 0.0010 \\
            Fe & 0.0042 \\
            Ga & 0.0005 \\
            Mg & 0.0007 \\
            Mn & $<0.0001$ \\
            Si & 0.0023 \\
            Ti & $<0.0002$ \\
            Zn & 0.0009 \\
            \hline
        \end{tabular}
    \end{table}
    \section{Total uncertainty on normalized \gammaray counts}\label{app:error:normalized:counts}

\begin{table}[t]
    \centering
    \begin{tabular}{c l}
    \multicolumn{2}{l}{\it Neutron rate ($n\pm\delta_n$)}\\
    $\delta_N^S$ : & Poisson statistics on neutron counts.\\
    $\delta_n^\Omega$ : & Uncertainty on solid angle subtended by neutron monitor.\\
    $\delta_n^t$ : & Uncertainty on the time of integration.\\
    \multicolumn{2}{l}{\it Foil counting ($C\pm\delta_C$)}\\
    $\delta_C^S$ : & Poisson statistics on \gammaray counts.\\
    $\delta_C^\Omega$ : & Uncertainty on solid angle subtended by \gammaray detector.\\
    $\delta_C^F$ : & Error on the parameters of the fit (assuming Gaussian peak).\\
    $\delta_C^P$ : & Error due to arbitrary choice of fitting meta parameters.\\
    $\delta_C^T$ : & Uncertainty on the time of integration.\\
    $\delta_C^{\Delta t}$ : & Uncertainty on the time $\Delta t$ between irradiation and counting.
    \end{tabular}
    \caption{List of the possible source of errors in final results.}
    \label{tab:error:source}
\end{table}

We want to evaluate the absolute uncertainty on the \gammaray counts normalized to the neutron rate ($C_n$), which we will refer to as $\delta_{Cn}$. We can assume to have the sources of error reported in~\cref{tab:error:source}, and we can combine them in two steps. First we decompose the error into neutrons and \gammaray:
\begin{eqnarray}
    C_n &=& \frac{C}{n} \\
    \left(\frac{\delta_{Cn}}{C_n}\right)^2 &=& 
        \frac{1}{C_n^2}\left[\left(\frac{\partial C_n}{\partial C} \delta_C \right)^2
        + \left(\frac{\partial C_n}{\partial n} \delta_n \right)^2\right]\nonumber\\
        &=& \left(\frac{\delta_C}{C}\right)^2+\left(\frac{\delta_n}{n}\right)^2
\end{eqnarray}
where $C$ is the total number of \gammaray counts, and $n$ is the neutron rate. Then we can focus on combining the sources of error in~\cref{tab:error:source} into $\delta_n$ and $\delta_C$.

\subsection{Uncertainties on neutron rate}\label{sec:error:neutrons}
Starting from the measured neutron rate, $n = \frac{N}{\Delta t}$, we assume the only sources of error can be the Poisson statistics on the total number of neutron counts $N$ measured during irradiation, the uncertainty on the neutron monitor position during measurements and the time of integration $t$. The neutron rate can also be calculated as $n = r \Omega \epsilon$ for a isotropic source: the total neutron production rate of the generator $r = \frac{N_0}{\Delta t}$, the solid angle subtended by the detector $\Omega$ and the efficiency of the detector $\epsilon$. The relative error on the neutron rate, then, can be calculated as:
\begin{eqnarray}
    \left(\frac{\delta_n}{n}\right)^2 &=& 
        \left(\frac{\delta_N^S}{N} \right)^2
        + \left(\frac{\delta_n^\Omega}{\Omega}\right)^2
        + \left(\frac{\delta_n^{t}}{t}\right)^2.
\end{eqnarray}
The Poissonian uncertainty on the total number of neutron counts will be equal to $\sqrt{N}$. $\Delta t$ is typically $1$ hr or longer with an uncertainty of about $10$ s, meaning $\frac{\delta_n^{t}}{t}<0.3 \%$ can be neglected. 

To estimate the error on the solid angle $\Omega$, we can simplify our geometry as if the neutron source was a point. The solid angle subtended by a cone having half opening angle $\theta$ is:
\begin{equation}
    \Omega = 4 \pi \sin^2\left(\frac{\theta}{2}\right).
\end{equation}
We placed the neutron monitor $R=1$ m away from the neutron source, and the detector diameter is $D=5$ cm. We can assume $\theta$ and $\theta \approx \frac{D}{2R}\ll1$, so that $\sin^2\left(\frac{\theta}{2}\right) \approx \frac{\theta}{2}$ and the solid angle becomes:
\begin{eqnarray}
    \Omega \approx 4 \pi \left(\frac{\theta}{2}\right)^2 = 
       \frac{\pi D^2}{4R^2}.
\end{eqnarray}
From this formula we can calculate the error on $\Omega$, considering $R$ to be the only source of uncertainty:
\begin{equation}
    \frac{\delta_n^\Omega}{\Omega}
    = \frac{2\delta_R}{R}.
\end{equation}
If we place the detector at $R = 100$ cm, with a precision on the position of $\delta_R = 1$ cm, then this relative uncertainty is worth $2$\%.

\subsection{Uncertainties on \gammaray counts}\label{sec:error:gamma}
As usual, the Poisson statistics can be estimated as:
\begin{equation}
    \delta_C^S = \sqrt{C}
\end{equation}
this uncertainty comes from the fact that this is a counting experiment. 

Since the number of counts under the peak comes from a fitting, we need to consider also the uncertainties coming from the fitting routine itself. The first part of this uncertainty comes from the confidence on the parameters of the function used to perform the fit. We are using a Gaussian fitting, for which the fitting function is:
\begin{equation}
    f(E) = A e^{-\frac{1}{2}\left(\frac{E-\mu}{\sigma}\right)^2}
\end{equation}
where $E$ is the energy axis of the spectrum measured by the detector, $\mu$ is the mean energy of the peak, $\sigma$ is a measure of the resolution of the detector and $A$ is the amplitude of the peak. The total number of counts is equal to the integral of this function over the energy interval of the fit $[E_0, E_1]$. Typically $f(E_0) \approx f(E_1) \approx 0$, then:
\begin{eqnarray}
    C &=& \int_{E_0}^{E_1} A e^{-\frac{1}{2}\left(\frac{E-\mu} 
    {\sigma}\right)^2}\, dE
    \approx \int_{-\infty}^\infty A e^{-\frac{1}{2}\left(\frac{E-\mu} 
    {\sigma}\right)^2}\, dE\nonumber\\
    &=& A\sigma \sqrt{2\pi}.
\end{eqnarray}
Then the error on $C$ due to the fitting uncertainties of $A$ and $\sigma$ is:
\begin{eqnarray}
    \left(\frac{\delta_C^F}{C} \right)^2
    &=& \left(\frac{\delta_\sigma}{\sigma}\right)^2+\left(\frac{\delta_A}{A}\right)^2.
\end{eqnarray}
The fitting procedure induces another uncertainty because we can arbitrarily choose the energy region $[E_0, E_1]$ over which we want to perform the fitting. Different energy regions give fits that accurately reconstruct the experimental fit within the counting statistics per bin of the measured spectrum. We estimated this uncertainty by manually varying the interval of the fit, finding that:
\begin{equation}
    \frac{\delta_C^P}{C} \approx 5 \%.
\end{equation}

\begin{figure}[t]
    \centering
    \includegraphics[width=0.6\linewidth]{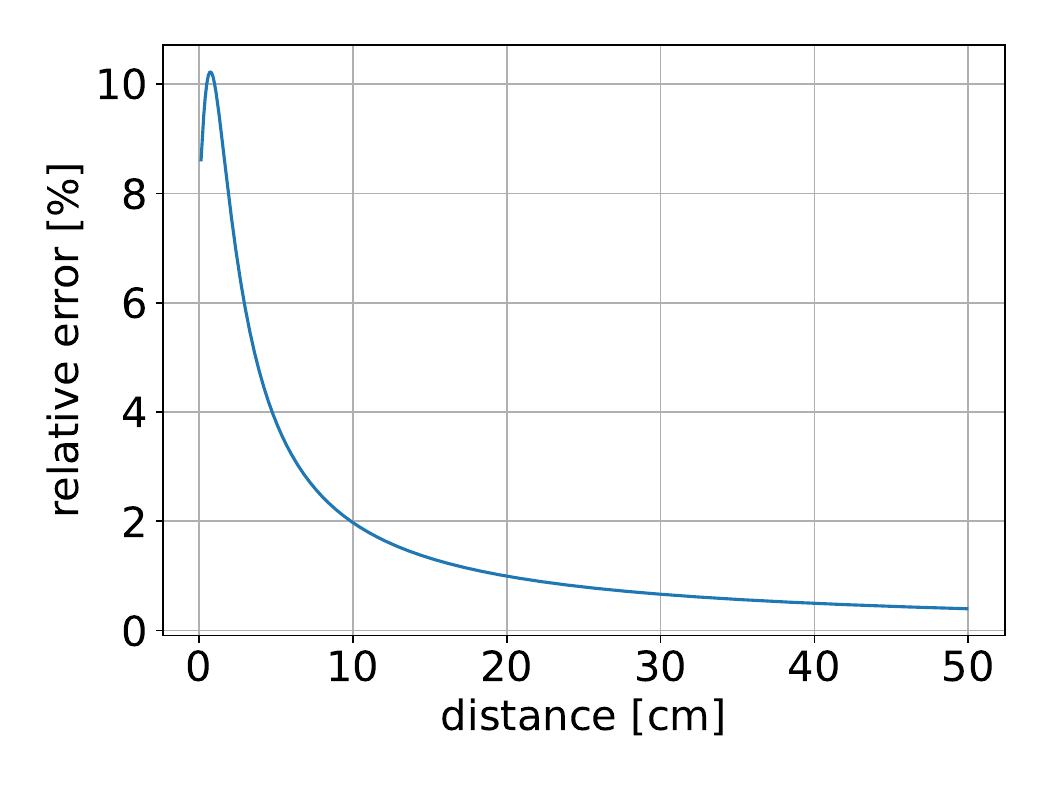}
    \caption{Relative error on \gammaray counts due to the uncertainty on the solid angle subtended by the detector.}
    \label{fig:uncertainty}
\end{figure}

Finally we need to calculate the uncertainty on the solid angle subtended by the \gammaray detector during foil counting. Since the foil is much smaller than the detector, we simplify the foil to be a point source. We follow a procedure similar to the one described in~\cref{sec:error:neutrons}. In this case, we want to scope all possible positions of the foil, from the case in which the capsule is in touch with the detector, to the case in which the rabbit is placed far away. Then we can calculate the angle that defines the cone of the field of view as:
\begin{equation}
    \theta = \arccos\left(\frac{R}{\sqrt{R^2+h^2}}\right),
\end{equation}
where $h$ is the radius of the detector base. The solid angle becomes:
\begin{equation}
    \Omega = 2\pi \left(1-\cos(\theta)\right) = 2\pi \left(1-\frac{R}{\sqrt{R^2 + h^2}} \right). 
\end{equation}
As before, we assume $R$ is the only source of error in this formula:
\begin{eqnarray}
    \delta_\Omega &=& \sqrt{\left(\frac{\partial \Omega}{\partial R} \delta_R \right)^2} \nonumber\\
    &=& \frac{2\pi h^2}{\left(R^2 + h^2\right)^\frac{3}{2}} \delta_R
\end{eqnarray}
and we finally get the relative error due to the solid angle definition:
\begin{equation}
    \frac{\delta_C^\Omega}{C} 
    = \frac{\delta_\Omega}{\Omega}
    = \frac{h^2}{\left(R^2+h^2\right)\left(\sqrt{R^2+h^2}-R\right)} \delta_R.
\end{equation}
In our experiments $h = 1.27$ cm and $\delta_R = 1$ mm, then the error behaves as in~\cref{fig:uncertainty}. In~\cref{tab:error:source} we report few values for a source positioned in contact with the detector, one short capsule away from the detector and one long capsule from the detector. These are the positions of the \ce{^{137}Cs} source used in~\cref{sec:gamma:attenuation}.

It is worth noting that the time of integration for \gammaray counting was longer than $1$ hr, with an uncertainty of $0.1$ s. Similarly we can estimate the error due to the uncertainty on the time $\Delta t$ between the irradiation and the counting phase. We tried to keep $\Delta t \approx 2$ min in all experiments, however both the neutron generator and the \gammaray spectrometers had to be operated manually. By consulting the log of the digitization files, we noticed that $\Delta t$ has a variability of $7$ s. The exponential radioactive decay law tells us that the total number of \gammaray emitted in a time interval $T$, starting $\Delta t$ after the end of irradiation is:
\begin{equation}
    N_\gamma = N_0 e^{-\lambda \, \Delta t} \left(1-e^{-\lambda T} \right),
\end{equation}
where $N_0$ is the number of unstable nuclei at the end of the irradiation and $\lambda = \log(2)/t_\frac{1}{2}$ is the corresponding decay constant, which is related to the nuclei half-life $t_\frac{1}{2}$. The uncertainty on $N_\gamma$ then:
\begin{equation}
    \left(\frac{\delta N_\gamma}{N_\gamma} \right)^2 = \frac{\log(2)}{t_\frac{1}{2}} \left(\delta_{\Delta t}^2 + \delta_T^2 \right).
\end{equation}
The relative error is higher for short half-life decays. In our experiments the shortest half-life was $9.5$ min, which corresponds to a relative error of $0.9$ \%.

To combine all these uncertainties together, we follow a similar procedure as in~\cref{sec:error:neutrons}:
\begin{eqnarray}
    \left(\frac{\delta_C}{C}\right)^2 = 
        \frac{{\delta_C^S}^2+{\delta_C^F}^2+{\delta_C^P}^2}{C^2}
        + \left(\frac{\delta_C^\Omega}{\Omega}\right)^2
        + \left(\frac{\log(2)}{t_\frac{1}{2}}\right)^2 \left({\delta_C^{T}}^2+{\delta_C^{\Delta t}}^2\right).
\end{eqnarray}

\begin{table}[t]
    \centering
    \begin{tabular}{l r}
    Distance [cm] & $\frac{\delta_\Omega}{\Omega}$\\
    \midrule
    $0.33$ &  $9.5$ \%\\
    $2.40$ &  $6.9$ \%\\
    $4.95$ &  $3.9$ \%
    \end{tabular}
    \caption{Relative uncertainty on the solid angle as a function of the distance between the \ce{^{137}Cs} \gammaray source. We consider the three distances used in the experiments conducted to assess the \gammaray attenuation of the capsule.}
    \label{tab:error:results}
\end{table}
%\end{multicols}

\bibliography{references}
%\printbibliography

\end{document}